\documentclass[twocolumn,english,prd,superscriptaddress,nofootinbib,notitlepage]{revtex4-1}
\pdfoutput=1

\usepackage[latin1]{inputenc}
\usepackage{babel}
\usepackage{url}

\usepackage{ifpdf}
\ifpdf   
    \usepackage[pdftex]{graphicx}
    \usepackage[hyperfootnotes=false,pdfstartview=,colorlinks=true,linkcolor=purple,citecolor=blue,urlcolor=blue,bookmarks=false]{hyperref}    \DeclareGraphicsExtensions{.png,.pdf}
    \usepackage{pgf}
\else    
    \usepackage[usenames,dvipsnames]{color}
    \usepackage[hyperfootnotes=false,pdfstartview=,colorlinks=true,linkcolor=BrickRed,citecolor=blue,urlcolor=blue,bookmarks=false]{hyperref}
    \usepackage{graphicx}
    \DeclareGraphicsExtensions{.eps}   
\fi

\usepackage{amssymb,amsmath}

\newcommand{\dd}{\textrm{d}}

\newcommand{\alm}{$a_{\ell m}$}
\newcommand{\unit}[1]{\ensuremath{\, \mathrm{#1}}}

\begin{document}

\title{Measuring our Peculiar Velocity by ``Pre-deboosting'' the CMB}

\author{Alessio Notari}
\affiliation{Departament de F\'isica Fonamental i Institut de Ci\`encies del Cosmos, Universitat de Barcelona, Mart\'i i Franqu\`es 1, 08028 Barcelona, Spain}

\author{Miguel Quartin}
\affiliation{Instituto de Física, Universidade Federal do Rio de Janeiro, CEP 21941-972, Rio de Janeiro, RJ, Brazil}
\affiliation{Institute of Theoretical Astrophysics University of Oslo, 0315 Oslo, Norway}

\date{\today}

\begin{abstract}
    It was recently shown that our peculiar velocity $\boldsymbol{\beta}$ with respect to the CMB induces mixing among multipoles and off-diagonal correlations at all scales which can be used as a measurement of $\boldsymbol{\beta}$, which is independent of the standard measurement using the CMB temperature dipole. The proposed techniques rely however on a perturbative expansion which breaks down for $\,\ell \gtrsim 1/\beta \approx 800$. Here we propose a technique which consists of deboosting the CMB temperature in the time-ordered data and show that it extends the validity of the perturbation analysis multipoles  up to $\,\ell \sim 10000$. We also obtain accurate fitting functions for the mixing between multipoles valid in a full non-linear treatment. Finally we forecast the achievable precision with which these correlations can be measured in a number of current and future CMB missions. We show that Planck could measure the velocity with a precision of around $60$ km/s, ACTPol in 4 years  around $40$ km/s, while proposed future experiments could further shrink this error bar by over a factor of around 2.
\end{abstract}

\maketitle

\section{Introduction}\label{sec:intro}

The dipole of the CMB is measured to be much larger than the other multipoles and this is usually attributed to a Doppler effect due to our peculiar velocity $\boldsymbol{\beta}$ with respect to the CMB rest frame. Under this assumption we can infer, by combining the measured temperatures of WMAP dipole~\cite{Hinshaw:2008kr} with the COBE monopole~\cite{Lineweaver:1996xa,Mather:1998gm}, its direction to be $\, l = 263.99^\circ ± 0.14^\circ$, $b= 48.26^\circ ± 0.03^\circ\,$ in galactic coordinates, and its modulus to be $\beta \equiv |\boldsymbol{\beta}| = (1.231 \pm 0.003) \times 10^{-3}$.
These very precise numbers rely however entirely on the above assumption, but generically the CMB dipole is not necessarily due only to a relative velocity. One way to test this assumption was proposed in~\cite{Burles:2006xf} where it was shown that our peculiar velocity could also be measured using the asymmetry in the location of the peaks of the power spectrum between forward and backward hemispheres, achieving a possible detection of $\beta$ at $2 - 3 \sigma$ for resolution $\ell = 1000 - 1500$. Recently it has also been pointed out~\cite{Challinor:2002zh,Kosowsky:2010jm,Amendola:2010ty} that {\it all} the CMB multipoles $a_{\ell m}$ have a correction due to our local peculiar velocity because the primordial anisotropies are distorted by the Doppler and aberration effects.  This shows up as a correlation between different multipoles $\ell$ and  could be used as an alternative way of measuring our velocity, as an independent consistency check. This fact might offer also a way to test the isotropy of the Universe on very large scales: it is known in fact that the CMB sky and other observations seem to exhibit some anomalies on the very large scales and this effect offers an observational handle which could either confirm and make more robust the standard assumptions or perhaps point to global anisotropies of the Universe.

The analysis of these CMB correlations has been performed~\cite{Challinor:2002zh,Kosowsky:2010jm,Pereira:2010dn,Amendola:2010ty} relying upon a Taylor expansion in orders of $\beta$ of boost effects (Doppler and aberration) on the multipole coefficients $a^{X}_{\ell m}$, where $X$ stands for $T$ (temperature) or $E, B$ (the $E$ and $B$ modes of polarization). It turns out, however, that each of the series coefficients brings together ever higher powers of $\,\ell$, effectively transforming the expansion into one of powers of the product $\,\beta\,\ell$. Since $\beta = 1.231 \times 10^{-3}$ this means that for $\ell \gtrsim 800$ the series can no longer be relied upon. We shall therefore distinguish between two regimes of angular scales where the the above correlations show up. In the first regime on large scales, $\ell \ll 1/\beta \simeq 800$, we can use a perturbative approach and therefore precisely predict that there are  correlations of ${\cal O} (\beta \ell)$ between neighbor $\ell$'s and look for a signal in the CMB in the form of correlators $\,a_{\ell m}^* a_{\ell+1 m'}$.  Moreover it was shown in~\cite{Amendola:2010ty} that one can non-trivially and very efficiently define three different estimators (for $m'-m=0,\pm 1$) in order to measure directly the three cartesian components of $\boldsymbol{\beta}$, without having to scan all the possible directions in the sky (i.e., without the need to compute and minimize a numerical $\chi^2$ for the correlations for each direction in a grid of $\{\theta,\phi\}$ coordinates).

The multipoles in the complementary regime ($\ell \gtrsim 1/\beta$) cannot be so easily treated but nevertheless also carry information about our peculiar velocity in the form of similar correlations, and it turns out that such information is needed in order to reach a signal-to-noise ratio larger than 1. In this regime the deviation angle $\theta-\theta'$ due to aberration is larger than the angular scale of interest $1/\ell$, so things are much more complicated. There are in fact correlations which are large (${\cal O} (1)$!)  and nonzero  also for distant $\ell$'s given by a transformation $ a^{[A]}_{\ell m} \;=\; \sum_{\ell'=0}^\infty K_{\ell' \, \ell\, m}\, a^{[P]}_{\ell' m}\,,$ between the aberrated $[A]$ frame and the primordial $[P]$ correlations. One in principle would have to compute the matrix elements $\,K_{\ell' \ell m}\,$, sometimes referred to as the \emph{aberration kernel}, which are integrals of spherical harmonics with different arguments, plug the matrix elements into all the possible two-point correlation functions and compare with the data. In fact, as we will show in Section~\ref{sec:nonlinear} the correlations in non-neighboring multipoles ($a_{\ell m}^* a_{\ell\pm n\, m'}$, $n>1$) also carry a measurable signal, which should be taken into account to measure the velocity with reasonable precision. While in principle straightforward this procedure has some disadvantages: \emph{(i)} this can be a heavy and delicate numerical task, because of the highly oscillating integrands and also because of the huge number of correlators that one would have to consider (future experiments propose to measure all multipoles as far as $\ell \sim 3000+$ for both temperature and polarization); \emph{(ii)} it is not obvious to understand in this case whether the three simple estimators can be written explicitly for the three cartesian components of the velocity for any $n$ in $a_{\ell m}^* a_{\ell\pm n\, m'}$, so perhaps the procedure would have to be carried out scanning the sky in all possible directions which would probably make this approach even more expensive in terms of computational time.

Even though we explore this approach further and propose a solution (in Section~\ref{sec:nonlinear}) to the first of the two challenges  listed above, the main point of this paper is to suggest a very simple trick which may be used to overcome all of these problems and measure directly the three components of $\boldsymbol{\beta}$ from a map in a much faster and straightforward way. Using this result we are able to predict with which precision we can measure $\boldsymbol{\beta}$ for several future experiments. This trick relies on the fact that we may already {\it assume} to know the direction and modulus of the velocity $\boldsymbol{\beta}$ from the CMB dipole and that we can the use the other multipoles with their correlations as an independent consistency check, in order to confirm (or not) the assumed value of $\boldsymbol{\beta}$ up to some precision.

The trick (described in more detail in Section~\ref{sec:pre-deboost}) is as follows. Given the central assumed value of the velocity $\,\boldsymbol{\beta}_{\rm dip} = \boldsymbol{\beta}_{\rm dip}^{\rm fit} \pm \delta\boldsymbol{\beta}_{\rm dip}$  indicated by the dipole, we may take the CMB map $T(\theta,\phi)$ and ``deboost'' it by a Lorentz transformation  into the frame with opposite velocity $\,-\boldsymbol{\beta}_{\rm dip}^{\rm fit}$. As discussed in~\cite{Menzies:2004vr} such transformation should be carried in the CMB time-ordered data (TOD), i.e., before data treatment to extract the harmonic multipole coefficients \alm\ and thus before one constructs the CMB temperature maps. Thus we call this technique ``pre-deboosting'' the CMB data. In the new (pre-deboosted) frame the residual velocity  $\,\boldsymbol{\beta}_{\rm res}$ compared to the CMB frame is expected to be given just by the residual error $\,\delta\boldsymbol{\beta}_{\rm dip}\,$ on the experimental determination of $\,\boldsymbol{\beta}_{\rm dip}$. As quoted above, such error is currently approximately $\,|\delta\boldsymbol{\beta}_{\rm dip}|  \,\simeq\, 3\,\times\,10^{-6}$. As a consequence the correlations due to Doppler and aberration in this new frame are expected to be of the order $\,{\cal O}(\beta_{\rm res} \; \ell)\,$ for $\,\ell \ll 1/(\beta_{res})\simeq 3\times10^5\,$ and so we could safely use the first order perturbative equations to compute the correlation functions $\,a^*_{\ell m} a_{\ell+1 m'}\,$ up to, say, $\,\ell \simeq 10000\,$ which is more than enough for all future experiments.
Note that this allows us to use directly again the three efficient define estimators (for $m'-m=0,\pm 1$) for measuring direction and modulus of the residual velocity.

An unexpected but very interesting prospect is the case in which the measured velocity turns out to be different from the expected $\beta_{\rm res}$. This would be a clear indication that the CMB dipole is not completely (nor around $99\%$, as sometimes stated) due to our peculiar velocity alone.
Therefore, this would imply that there are other contributions to the CMB dipole, and it would be interesting to understand whether such correlations may distinguish even the nature of such contributions: adiabatic, isocurvature perturbations, dipolar lensing or other more exotic contributions.

The ability to measure exotic contributions to the dipole is of great interest to test Cosmology on very large scales, which could hide non-trivial phenomena, as suggested by some reported anomalies on the low-$\ell$ CMB multipoles itself~\cite{Copi:2010na, Bennett:2010jb}.
For instance, it could provide valuable information about some proposed tilted cosmological models in which the dipole arises partly due to primordial superhorizon-scale isocurvature fluctuations~\cite{Turner:1991dn,Grishchuk:1992wv,Langlois:1995ca,Erickcek:2008sm,Zibin:2008fe},
which could provide a possible explanation some recent controversial claims of a high galaxy cluster~\cite{Kashlinsky:2008ut,Kashlinsky:2009dw,AtrioBarandela:2010wy} and galaxy~\cite{Watkins:2008hf} bulk flow on large scales and it could be used as a test of non-standard cosmological models. In any case, if $\beta_{\rm res}\neq 0$ we would also be able to measure the direction of our velocity with an error $\delta\theta=\delta\beta/\beta_{\rm res}$ (see~\cite{Amendola:2010ty}).

This paper is organized as follows. We start by discussing in Section~\ref{sec:nonlinear} the full non-linear approach to estimate our velocity, and derive a very accurate fitting function for the oscillating integrals. We then discuss in Section~\ref{sec:pre-deboost} the pre-deboost technique originally proposed by~\cite{Menzies:2004vr}. In Section~\ref{sec:applications} we forecast the sensitivity expected from present and future experiments.
Finally, we draw our conclusions and summarize our results.

\section{The Full Non-linear Fit Technique}\label{sec:nonlinear}

\subsection{Fitting functions for the aberration kernel}

It was shown in~\cite{Challinor:2002zh,Kosowsky:2010jm,Amendola:2010ty} that when subjected to an aberration effect the  \alm coefficients of the spherical harmonic decomposition of the temperature (and polarization) contrast transform in the following way
\begin{equation}\label{eq:aberrated-alm}
    a^{X\,[A]}_{\ell m} \;=\; \sum_{\ell'=0}^\infty K_{\ell' \, \ell\, m}^X \, a^{X\,[P]}_{\ell' m}\,,
\end{equation}
where the superscript $X$ stands for either temperature ($T$) or one of the two independent modes of polarization ($E$ and $B$) and where $[A]$ denotes the aberrated coefficients, to be contrasted with $[P]$, the primordial (non-aberrated) ones.

In the case of temperature, the exact coefficients of~\eqref{eq:aberrated-alm} are given by~\cite{Kosowsky:2010jm} (we here follow~\cite{Amendola:2010ty} for the convention of the sense of $\boldsymbol{\beta}$ which results in an overall sign change in the velocity $\beta$)
\begin{equation}\label{eq:non-linear-coef}
\begin{aligned}
     K_{\ell'\, \ell\, m}^T = \int_{-1}^{1}  \frac{\dd x}{\gamma\, (1-\beta x)} \,\tilde{P}_{\ell'}^m(x) \, \tilde{P}_{\ell}^m \! \left(\frac{x - \beta}{1 - \beta  x}\right) ,
\end{aligned}
\end{equation}
where $\,\gamma \equiv 1/\sqrt{1-\beta^2}\,$ is the standard Lorentz factor and
\begin{equation}\label{eq:non-linear-fit-m0}
    \tilde{P}_{\ell}^m(x) \;\equiv\; \sqrt{\frac{2\ell+1}{2} \frac{ (\ell-m)!}{(\ell+m)!}} \,P_{\ell}^m(x) \,,
\end{equation}
and where $P_{\ell}^m$ are the associated Legendre polynomials. For polarization the formulae are similar if one makes use of spin-weighted spherical harmonics\footnote{For a discussion on spin-weighted spherical harmonics see~\cite{Newman:1966ub,Zaldarriaga:1996xe}} (${}_s\tilde{Y}_{\ell}^m$). Following~\cite{Challinor:2002zh}, we get
\begin{equation}\label{eq:non-linear-coef-pol}
\begin{aligned}
     K_{\ell'\, \ell\, m}^P &\;=\; \frac{1}{2} \bigg[ {}_2K_{\ell'\, \ell\, m} + {}_{-2}K_{\ell'\, \ell\, m} \bigg] ,
\end{aligned}
\end{equation}
in which
\begin{equation}\label{eq:non-linear-coef-pol2}
\begin{aligned}
     {}_sK_{\ell'\, \ell\, m}^P = \int_{-1}^{1}  \frac{\dd x}{\gamma\, [1-\beta x]} \,{}_s\tilde{P}_{\ell'}^m(x) \, {}_s\tilde{P}_{\ell}^m \! \left(\frac{x - \beta}{1 - \beta  x}\right) \,
\end{aligned}
\end{equation}
and where in turn $\,{}_s \tilde{P}_{\ell}^m(x) \equiv \sqrt{2\pi}\;{}_s Y_{\ell}^m(x,\phi=0)$, the spin-weighted spherical harmonics evaluated at $\phi=0$, which can be written as~\cite{durrer2008cosmic}
\begin{equation}\label{eq:spin-weighted-P}
\begin{aligned}
    {}_s\tilde{P}_{\ell}^m & (x) \;=\; \sqrt{{2\ell+1\over 2}{(\ell+m)!(\ell-m)!\over (\ell+s)!(\ell-s)!}
    } \left(\frac{1-x}{2}\right)^\ell  \\
    \times \sum_r & {\ell- s \choose r}{\ell+s \choose r+s-m}
    (-1)^{\ell-r-s+m}\sqrt{\frac{1+x}{1-x}}^{2r+s-m} \!,
\end{aligned}
\end{equation}
where the sum in $r$ is to be carried out in the range max$\{0,\,m-s\} \le r \le \mbox{min}\{\ell-s,\,\ell +m\}$. Note that~\eqref{eq:non-linear-coef-pol2} and~\eqref{eq:spin-weighted-P} reduce respectively to~\eqref{eq:non-linear-coef} and~\eqref{eq:non-linear-fit-m0} for $s=0$.

The oscillatory nature of these integrals poses a numerical challenge, which make their direct computation very slow for high $\ell$. In a recent work~\cite{Chluba:2011zh} a recursive method was developed which is claimed to allow fast and accurate evaluation of these integrals. In this section we follow instead a different route and compute some of the integrals just by numerical integration (see Appendix~\ref{app:details-numerical} for more details). Surprisingly, we found out numerically that these integrals can be fit \emph{very precisely }by Bessel functions, which greatly simplifies the analysis.

Following~\cite{Challinor:2002zh} it is convenient to define the quantities
\begin{align}\label{eq:challinor-fun}
    {}_sG_{\ell\, m} \;\equiv\; \sqrt{\frac{\ell^2-m^2}{4\ell^2-1}\left[1-\frac{s^2}{\ell^2}\right]},
\end{align}
where again $s$ represents a spin weight which is $\,0\,$ for temperature and $\,2\,$ for the $E$ and $B$ modes of polarization. Note that we always have $\,0\leq{}_sG_{\ell\, m} \lesssim 1/2$, the lower limit being achieved when $|m|=\ell$ and the higher one when $m=0$. An exquisite fit for small scales to the full non-linear integral for general $m$ is given by
\begin{equation}\label{eq:non-linear-fit-n1}
\begin{aligned}
    K_{\ell-1\, \ell\, m}^T   &\;\simeq\; J_1\!\Big(\! -2\, \beta \,\ell \;{}_0G_{\ell\, m} \Big), \\
    K_{\ell+1\, \ell\, m}^T   &\;\simeq\; J_1\!\Big( \,2\, \beta \,(\ell+1)\;{}_0G_{\ell+1\, m} \Big),
\end{aligned}
\end{equation}
where $J_1$ is the Bessel function of the first kind. Moreover, we find that similar relations to the above one applies also for non-neighboring correlations (i.e., between any $\ell$ and $\ell \pm n$, $n\ge1$) and also for polarization:
\begin{equation}\label{eq:non-linear-fit-general}
\begin{aligned}
    K_{\ell-n\, \ell\, m}^X   &\;\simeq\; J_n\!\left(\!-2\, \beta \left[\prod_{k=0}^{n-1} \big[(\ell-k) \;{}_sG_{\ell-k\, m} \big]\right]^{1/n} \right), \\
    K_{\ell+n\, \ell\, m}^X   &\;\simeq\; J_n\!\left(\,2\, \beta \left[\prod_{k=1}^n \big[(\ell+k) \;{}_sG_{\ell+k\, m} \big]\right]^{1/n} \right),
\end{aligned}
\end{equation}
which we find to be accurate to around $\,0.2\%\,$ for all values of $\,\ell\,$ and $\,m\,$ and all values of $\,n$.\footnote{We explicitly checked the above fits for $\,n=\{0,\,1,\,2,\,3,\,4\}\,$ and $\,\ell \le 700\,$ for the case of temperature correlations. For polarization we only tested explicitly $\,n=\{0,\,1,\,2\}\,$ and $\,\ell \le 100$. Nevertheless, all results indicate that in both cases the precision would remain the same at higher $n$ and at smaller scales.} For $\,n=0\,$ one cannot apply~\eqref{eq:non-linear-fit-general} directly, but we find that an analogous fit is given by (with the same precision)
\begin{equation}\label{eq:non-linear-fit-n0}
\begin{aligned}
    K_{\ell\, \ell\, m}^X  \;\simeq\; J_0\!\Bigg(&  \beta\sqrt{2} \,\bigg[ \!-(\ell+1)\,(\ell+2)\,({}_sG_{\ell+1\, m})^2\,- \\
     &\ell\,(\ell-1) \,({}_sG_{\ell\, m})^2 + \ell(\ell+1) - \\
      & m^2 +1 -s^2 + \frac{s^2\,m^2}{\ell(\ell+1)} \bigg]^\frac{1}{2} \Bigg).
\end{aligned}
\end{equation}
As can be seen from the above relations, the temperature and polarization aberration kernels become almost indistinguishable for small scales (usually for any $\ell \gg 1$, or when $|m| \simeq \ell$, for $\ell \gg 5$).

A Taylor expansion of the Bessel functions yields to leading order
\begin{align}\label{eq:bessel-taylor}
    J_n(x) \;=\; \frac{1}{2^n \, n!} x^n\,+{\cal O}\big( x^{n+2} \big)\,,
\end{align}
which is valid for positive integer values of $n$. Expanding the above fits in orders of $\beta$ we find that the coefficients with the Doppler correction \emph{exactly} up to leading order for any value of $\,n\,$ for both temperature and polarization. In other words, for a given $n$ the fits are exact to order ${\cal O} \big(\beta ^ n \big)$. This was confirmed through direct analytic integration of the leading order of the Taylor expansion in $\beta$ of~\eqref{eq:non-linear-coef} for specific values of $\{\ell,\,\ell',\,m\}$. This is an interesting result, as the $\{\ell,\,\ell\pm3\}$, $\{\ell,\,\ell\pm4\}$ and so forth leading order coefficients were never derived before in the literature. The $\{\ell,\,\ell-3\}$ is for instance simply:
\begin{equation}\label{eq:n3-expansion}
\begin{aligned}
    K_{\ell-3\, \ell\, m}^T   =& -\frac{1}{6} \beta^3 \,\ell(\ell-1) (\ell-2) \;{}_0G_{\ell\, m} \;{}_0G_{\ell-1\, m} \;{}_0G_{\ell-2\, m} \\
    &+ {\cal O} \big( \beta^5 \big)\,,
\end{aligned}
\end{equation}
and similarly simple expressions hold for other $\{\ell,\,\ell\pm n\}$ correlations. Another cross-check of the above formulae is to confront the expansion up to second order with the coefficients in~\cite{Challinor:2002zh}, but there is a subtlety
involved and we come back to this issue in Section~\ref{subsec:temperature-coef}.

\begin{figure}[t]
    \includegraphics[width=8.2cm]{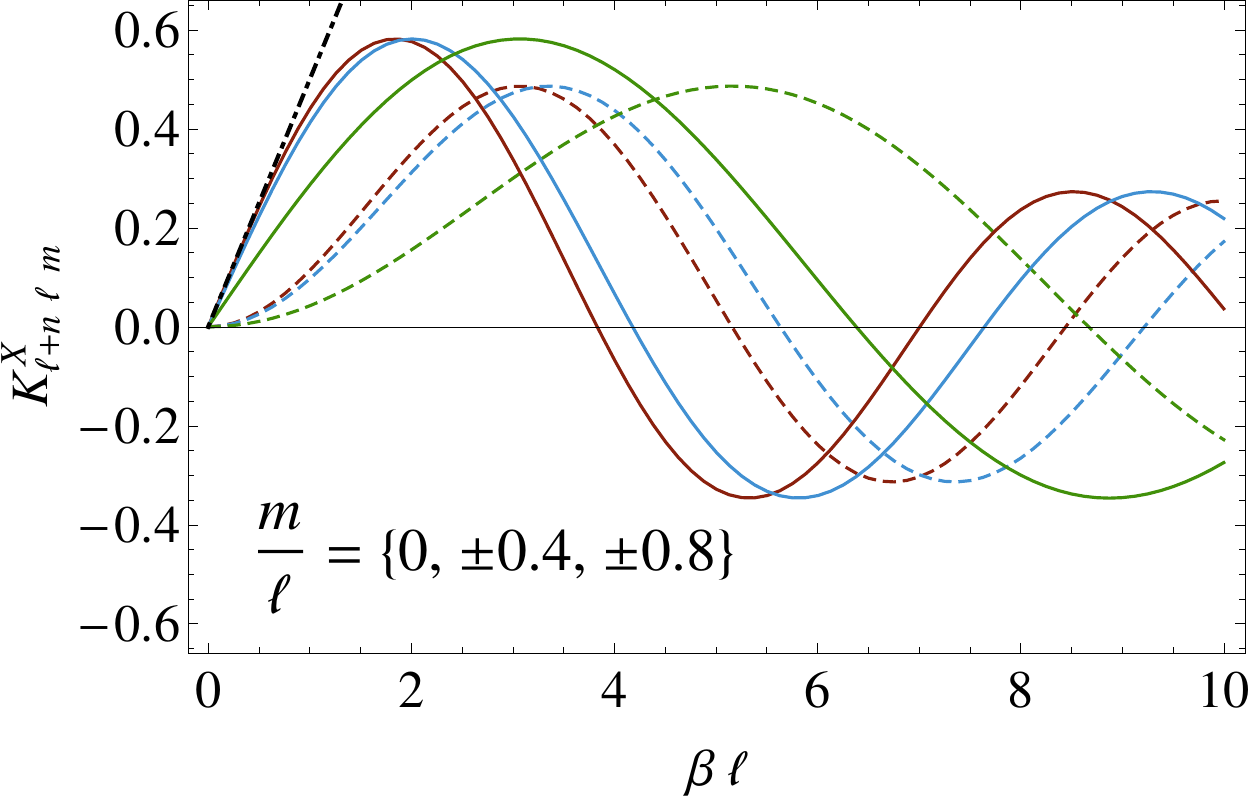}
    \caption{Some of the nonlinear coefficients in~\eqref{eq:non-linear-fit-general} for the cases $\,n=1\,$ (solid lines) and $\,n=2\,$ (dashed) and for different values of $m$ -- from left to right: $m=0$ (brown), $|m|/\ell=0.4$ (blue) and $|m|/\ell=0.8$ (green). As $|m|$ changes, the Bessel functions get stretched and therefore for a given $\ell$ the coefficients get smaller. Also plotted for comparison (black, dot-dashed) is the linear approximation for $m=0$. Note that for $\ell \gg 5$ the polarization and temperature coefficients become indistinguishable. }
    \label{fig:bessel}
\end{figure}

\begin{figure}[t]
    \includegraphics[width=8.4cm]{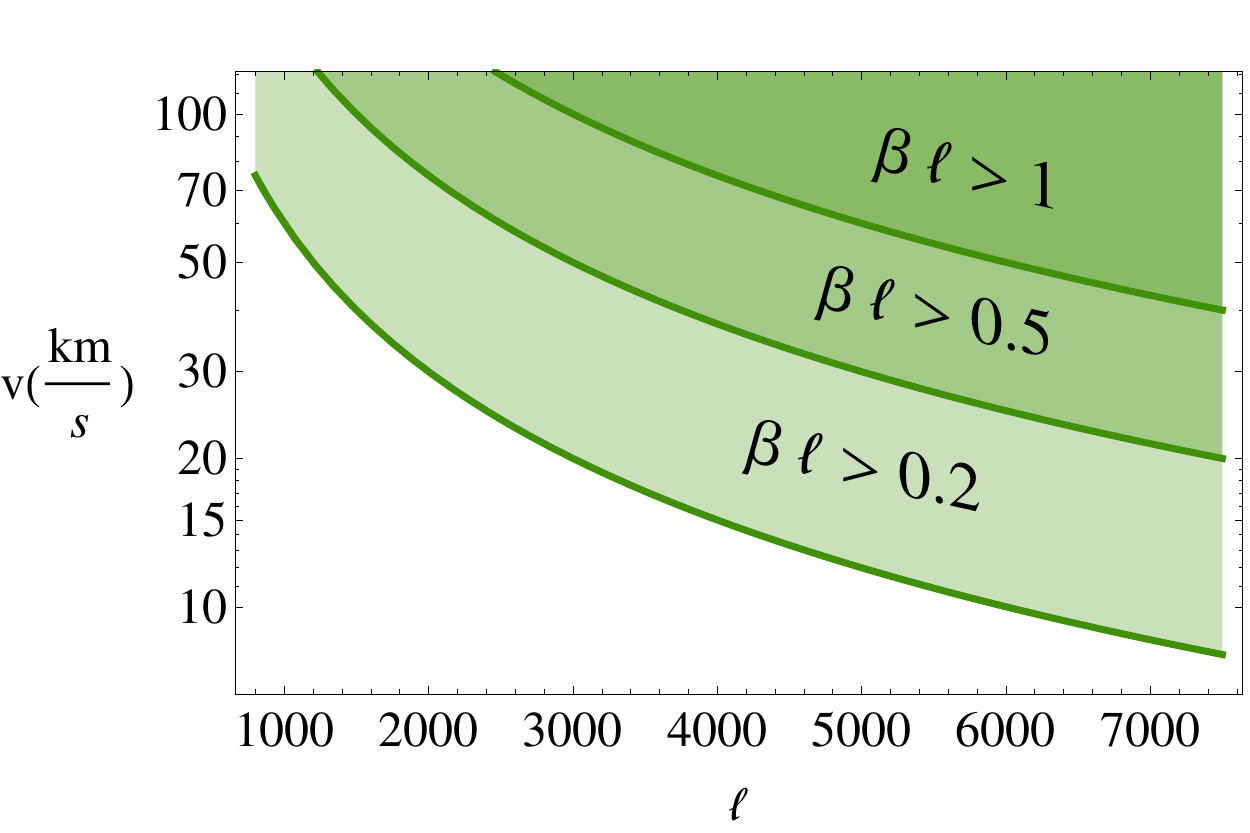}
    \caption{Regions for which the first order Taylor series used in Section~\ref{sec:applications} starts to break down, as a function of the aberration velocity.  The linear coefficients differs from the full result by roughly $\{0.4\%,\, 2\%,\,10\%\}$ for $\,\beta \ell = \{0.2,\, 0.5,\,1\}$. If one pre-deboosts the CMB, one expects $\,v_{\rm res} \ll 100$ km/s. }
    \label{fig:series-reliability}
\end{figure}

Figure~\ref{fig:bessel} depicts some of the nonlinear coefficients~\eqref{eq:non-linear-fit-general} for the cases $\,n=1\,$ and $\,n=2\,$ and for different values of $m$. As $m$ changes, the Bessel functions get stretched and therefore for a given $\ell$ the coefficients get smaller. This plot in particular was made computing~\eqref{eq:non-linear-fit-general} for $\ell = 100$, but as discussed above, the coefficients are essentially the same for any $\ell$, as long as $\,\beta \ll 1$. Figure~\ref{fig:series-reliability} depicts the regions for which $\beta \ell$ is greater than $0.2$, $0.5$ and $1$. Above these regions the linear coefficients differ from the non-linear ones by $\{0.4\%,\, 2\%,\,10\%\}$, respectively. Linear estimates are thus correct to within $10\%$ as long as the measured residual velocity lies within the $\beta \ell < 1$ region of this plot. This should always be the case if one pre-deboosts the CMB since then $\beta \rightarrow \beta_{\rm res} \ll 10^{-3}$, unless the residual contribution to the CMB dipole is unexpectedly large.

One could therefore conceive of the above fitting formula as  direct method to estimate the correlators which bypasses the need to pre-deboost the CMB. Nevertheless this is a subtle point since, as shown in~\cite{Amendola:2010ty}, in the velocity estimator there is a nearly-exact cancelation of the leading-order term in $\ell$ and it is thus possible that small corrections to the value of the coefficients lead to moderate corrections to the estimators. One would therefore need to carefully to check whether the accuracy provided by the fits here proposed is enough for such analysis. Moreover, as stated in Section~\ref{sec:intro} the usefulness of such method would also be dependent on whether also in this case  three simple estimators can be written explicitly for the three cartesian components of the velocity, thus avoiding the need to compute and minimize a $\chi^2$ for all possible sky directions. If both issues can be circumvented, then this would provide a technique to measure our velocity, which could be used as a cross-check for the pre-deboost technique.
In any case in the present paper we propose to use the easier route of using the pre-deboost technique, which is more straightforward and without complications, as we will show in section~\ref{sec:pre-deboost}. However, before doing that we analyze further the consequences of the approximations we have found using the Bessel functions.

As a side note, the fit~\eqref{eq:non-linear-fit-general} gets worse for high values of $\beta$ and in fact breaks down whenever $\,\beta$ is close to 1. This can be understood as one must have on physical grounds that $\,\beta < 1\,$ and from~\eqref{eq:non-linear-coef} that the integrals must approach zero in the limit $\,\beta \rightarrow 1$; the proposed fit instead does not go to zero in the same limit.\footnote{We found a correction to the proposed fits for $\,\beta \gtrsim 0.1$: the fits get much better as $\beta \rightarrow 1$ (but not exact) if one replaces $\,\beta\,$ by $\,\beta / (1-\beta^2)^{1/4}\,$ in the argument of the Bessel functions. Note that in practice, however, such a correction is  irrelevant as we know that $\beta \sim 10^{-3}$.} This fact is not important for the realistic case, since  $\beta\sim 10^{-3}$, but it might be relevant if one tries to use our fitting function to try and guess an exact analytic solution to the integrals~\eqref{eq:non-linear-coef}. In fact, the very high precision and breadth of applicability of an arguably simple fitting function~\eqref{eq:non-linear-fit-general} (which holds for any $n$ and either temperature or polarization) hints to the existence of an analytical solution involving these Bessel functions, which perhaps can only be derived in some special limit. An analytical expression for~\eqref{eq:non-linear-coef} or~\eqref{eq:non-linear-coef-pol2}, if it exists, could shed some light into the mathematics of this problem and into the limits of validity of the approximation needed to make such a derivation. Nevertheless, such analysis is beyond the scope of this work.


\subsection{Consequences of the exact coefficients}\label{subsec:temperature-coef}

The aberration kernel was computed up to second order in~\cite{Challinor:2002zh} and it is interesting to compare their results with ours. In fact, those analytic perturbative results motivated in part the fits~\eqref{eq:non-linear-fit-general} and~\eqref{eq:non-linear-fit-n0}. Note however that the coefficients in~\cite{Challinor:2002zh} differ slightly from the ones here and in~\cite{Kosowsky:2010jm,Amendola:2010ty} due to the fact that theirs were obtained assuming a transformation in total intensity and not temperature (as it should be done) and this entices a small correction due to the Doppler effect. In practice, in~\cite{Challinor:2002zh} the integrand in~\eqref{eq:non-linear-coef} and~\eqref{eq:non-linear-coef-pol2} had an incorrect exponent to the factor $\gamma\, (1-\beta x)$; to wit $+2$ instead of $-1$, which was first realized by~\cite{Kosowsky:2010jm}. This correction has two interesting consequences.

First, it was shown in~\cite{Challinor:2002zh} that there should be a bias in the measured CMB \emph{intensity} power spectrum proportional to $4\beta^2$; here we find instead that for the \emph{temperature} power spectrum (the one which is physically of more interest) the correction is exactly \emph{zero} at ${\cal O}\big(\beta^2\big)$. Thus if there is such a bias, it must be at most ${\cal O}\big(\beta^4\big)$\footnote{Note however that these results are valid only for the $C_{\ell}$ for a full-sky observation, while any experiment has inevitably some mask, implying that pseudo $C_{\ell}$ must be used, which leads to a bias which is present already at ${\cal O}(\beta)$, as shown in \cite{Pereira:2010dn}. }. Another interesting result in~\cite{Challinor:2002zh} is that there should be small ${\cal O} \big( \beta / \ell \big)$ cross-correlations  $EB$ and $TB$ which are usually absent unless parity is violated by some exotic process. In other words, there should be an additional term in~\eqref{eq:aberrated-alm}, to wit
\begin{equation}\label{eq:E-B-cross-correlation}
\begin{aligned}
     K_{\ell'\, \ell\, m}^{EB} &\;=\; \frac{i}{2} \bigg[ {}_2K_{\ell'\, \ell\, m} - {}_{-2}K_{\ell'\, \ell\, m} \bigg] .
\end{aligned}
\end{equation}
Here we find instead,  by evaluating analytically~\eqref{eq:non-linear-coef-pol2} for specific values of $\{\ell,\,\ell',\,m\}$ that this cross-correlation is again exactly \emph{zero}, at least up to ${\cal O}\big(\beta^6\big)$. These null results make both effects even less relevant than previously thought.

\subsection{Correlations among non-neighboring multipoles}


The relations above allow us to estimate the amount of signal into the correlation functions $\langle a^{X\,[A]}_{\ell m}{}^* a^{X\,[A]}_{\ell+n, m} \rangle$ between non-neighboring multipoles $n>1$, using~\eqref{eq:non-linear-fit-general} and $n=0$, using~\eqref{eq:non-linear-fit-n0}. Since the amount of signal between these multipoles are directly proportional to the coefficients~\eqref{eq:non-linear-coef}, a simple (and rough) estimate of the amount of signal in an experiment able to measure temperature multipoles up to a scale $\ell_{\rm max}$ is given by the sum in quadrature of the all the coefficients up to $\ell_{\rm max}$. As shown in~\cite{Amendola:2010ty}, the amplitude of the coefficients is just part of the story and the smoothness of the angular power spectrum is also relevant, and in particular for $\,C_\ell = C_{\ell +1}\,$ there is a cancelation of the leading order term. Nevertheless, a comparison of the values of such sum in quadrature between $\ell$ and $\ell + n$, $n \neq 1$, and the sum in quadrature of the ones between $\ell$ and $\ell + 1$ should give a hint of the relative amount of signal between these non-neighboring  multipoles. Since there are $2\ell +1$ coefficients for each $\ell$, we evaluate the quantity
\begin{align}\label{eq:other-correlators}
    \frac{\int_{0}^{\ell_{\rm max}}\dd \ell' \; (2\ell' + 1)\,\left[J_n\left(2 \,\beta \,\ell'\; \overline{{}_0G_{\ell\, m}}\right)\right]^2}{\int_{0}^{\ell_{\rm max}}\dd  \ell' \;(2\ell' + 1)\, \left[J_1\left(2 \,\beta \,\ell'\; \overline{{}_0G_{\ell\, m}}\right)\right]^2}\,,
\end{align}
for $\,n>1\,$ or, in the case $\,n=0\,$, instead the quantity
\begin{align}\label{eq:other-correlators-n0}
    \frac{\int_{0}^{\ell_{\rm max}}\dd \ell' \; (2\ell' + 1)\,\left[1-J_0\left(2 \,\beta \,\ell'\; \overline{{}_0G_{\ell\, m}}\right)\right]^2}{\int_{0}^{\ell_{\rm max}}\dd  \ell' \;(2\ell' + 1)\, \left[J_1\left(2 \,\beta \,\ell'\; \overline{{}_0G_{\ell\, m}}\right)\right]^2}\,,
\end{align}
where we approximated a sum in $\ell$ by an integral and where $\, \overline{{}_0G_{\ell\, m}}\,$ is the effective $m$-averaged quantity defined as
\begin{align}\label{eq:average-m}
     \overline{{}_0G_{\ell\, m}}=\frac{1}{2\ell+1}\sum_{m=-\ell}^{\ell} \,{}_0G_{\ell\, m}\,\simeq 0.39\,,
\end{align}
which for $\,\ell \gtrsim 10\,$  effectively does not depend on $\ell$. As we will discuss in more detail in Section~\ref{sec:applications}, for  Planck $\,\beta\ell_{\rm max} \simeq 2.5\,$ whereas for some proposed future experiments $\,\beta\ell_{\rm max} \simeq 5$. The above estimate tells us that for $\,\beta\ell_{\rm max} \simeq 2.5\,$ ($\beta\ell_{\rm max} \simeq 5$) the total amount of signal is, in comparison with the signal in $\{\ell, \, \ell+1\}$:  around $40\%$ ($100\%$), for $\{\ell, \, \ell+2\}$; around $13\%$ ($70\%$), for $\{\ell, \, \ell+3\}$; around $90\%$ ($270\%$), for $\{\ell, \, \ell\}$. These numbers are clearly non-negligible, contrary to what was claimed in~\cite{Kosowsky:2010jm} but in agreement with results in~\cite{Chluba:2011zh}. In the case of the diagonal ($n=0$) correlations they seem to be particularly large, and in fact can be higher than the $\{\ell, \, \ell+1\}$ signal.

The reason why it was sometimes thought that only the $\{\ell, \, \ell+1\}$ case was measurable stems from the Taylor expansion of the Bessel functions: for small $\,\beta\,\ell\,$, the $\{\ell, \, \ell\}$ and $\{\ell, \, \ell+2\}$ cases are suppressed by a factor of $\beta$, and the $\{\ell, \, \ell+3\}$ by a factor $\beta^2$. But both current and future experiments will reach very small scales, way beyond the validity of such leading-order expansion, and the total signal in non-neighboring multipoles will seemingly be crucial.

Although one could argue that the above estimates are not refined enough to make such claims, a thorough analysis of the full correlations in non-neighboring multipoles could be quite complicated. One would have to follow a similar procedure as in~\cite{Amendola:2010ty} and derive the corresponding estimators for the velocity for at least $n=0$ and $n=2$, taking care to isolate all the independent correlations (for $n=2$, for instance, there is a contribution from applying twice the $n=1$ correlations).

The method we propose and discuss below, based on pre-deboosting the CMB, circumvents all the difficulties stressed in this Section.

\section{The Pre-deboost Technique}\label{sec:pre-deboost}

Let $\boldsymbol{\hat{n}}$ be the direction of incoming light rays in a rest frame $S$ at rest with respect to the CMB and $\boldsymbol{\hat{n}'}$ the observed (by our CMB instrument) direction of the same light ray in a reference frame $S'$ which moves with velocity $\boldsymbol{\beta}$ relative to $S$. The relation between the two directions is the aberration effect and is given by~\cite{Calvao:2004ma,Amendola:2010ty}
\begin{align}
    \boldsymbol{\hat{n}^{\prime}} \,=\,  \frac{\boldsymbol{\hat{n}} +  \left[\gamma\,\beta + (\gamma - 1) \big(\boldsymbol{\hat{n}}\cdot\boldsymbol{\hat{\beta}} \big)\right]\boldsymbol{\hat{\beta}}} {\gamma(1+\boldsymbol{\beta}\cdot\boldsymbol{\hat{n}})} \,,
\end{align}
where again $\,\gamma \equiv (1-\beta^2)^{-1/2}$. Besides aberration, there is also the Doppler effect, which relates the observed frequency $\nu^\prime$ to the emission frequency $\nu$:
\begin{equation}
    \nu^{\prime} \,=\, \nu\,\gamma\,\big(1+ \boldsymbol{\beta}\cdot\boldsymbol{\hat{n}}\big)\,.
\end{equation}
The rest-frame CMB temperature field $T(\boldsymbol{\hat{n}})$ is similarly aberrated and we thus observe
\begin{align}
    T^\prime(\boldsymbol{\hat{n}^{\prime}}) \,=\, \frac{T(\boldsymbol{\hat{n})}}{\gamma \big( 1 - \boldsymbol{\beta}\cdot\boldsymbol{\hat{n}^\prime} \big)} \,.
\end{align}
The above relation is obviously invertible, and one can recover the original field $T(\boldsymbol{\hat{n}})$ by applying a (second) Lorentz boost to $T^\prime(\boldsymbol{\hat{n}^{\prime}})$ with velocity $-\boldsymbol{\beta}$.

The idea here is to use as velocity the one given by the measured CMB temperature dipole. In this case, the dipole translates trivially into a velocity $\,\boldsymbol{\beta}_{\rm dip} = \boldsymbol{\beta}_{\rm dip}^{\rm fit} \pm \delta\boldsymbol{\beta}_{\rm dip}\,$ for which the current uncertainty $\delta\boldsymbol{\beta}_{\rm dip}$ is less than $0.3\%$. We thus obtain
\begin{align}\label{eq:deboosted-T}
    T(\boldsymbol{\hat{n}})_{\rm deboosted}  \,=\,  \frac{T^\prime(\boldsymbol{\hat{n}^\prime)}}{\gamma \big( 1 + \boldsymbol{\beta}_{\rm dip}^{\rm fit}\cdot\boldsymbol{\hat{n}} \big)} \,.
\end{align}
This deboosted (or ``deaberrated'', as it was originally called in~\cite{Menzies:2004vr}) temperature field should give rise to \alm's with aberration induced correlations given no longer by $\,\boldsymbol{\beta}_{\rm dip} \sim 10^{-3}\,$ but instead by a much smaller velocity, of order $\,\delta\boldsymbol{\beta}_{\rm dip} \sim 3\times10^{-6}$. This means that even if future CMB experiments are able to probe very small scales such as $\ell \sim 10000$, the product $\,\delta\boldsymbol{\beta}_{\rm dip} \,\ell\,$ would still be much smaller than unity, and one could safely analyze these correlations using only the first order Taylor expansion coefficients. Moreover, in this case only the $\{\ell,\,\ell+1\}$ correlations are relevant. Therefore, by pre-deboosting one can rely on the velocity estimator already derived in~\cite{Amendola:2010ty}, and one need not worry about non-linear corrections nor about building a more complex estimator which would take into account the other $\{\ell,\,\ell+n\}$ correlations.

\begin{table*}[t]
\begin{tabular}{|c|c|c|c|c|c|c|}
    \hline
    \textbf{Experiment} & $\,\#\,\nu $ bands$\,$  & $\,10^6 \sigma_T (\frac{\mu K}{K})\,$ & $\,10^6 \sigma_{P}\,(\frac{\mu K}{K})\,$ & $\;\theta_{\rm fwhm}\;$ & $\;\;\;f_{\rm sky}\;\;\;$ & $\;\;S/N\;\;$ \\
    \hline \hline
    ACBAR '08 \cite{Reichardt:2008ay} & 1 & $0.9$ & -- & $4.8'$ & 1.7\% & \textbf{\textcolor[rgb]{0.8,0.00,0.00}{1.0}} \\
    \hline
    WMAP (9 years) \cite{Larson:2010gs,Collaboration:2011ck} & 5 & $14$ & $20$ & $13.2' - 52.8'$ & 78\% & \textbf{\textcolor[rgb]{0.8,0.00,0.00}{0.7}} \\
    \hline
    EBEX \cite{ReichbornKjennerud:2010ja} & 3 & 0.33 & 0.48 & $8'$ & 1\% & \textbf{\textcolor[rgb]{0.8,0.00,0.00}{0.9}} \\
    \hline
    BICEP2 (2 years)~\cite{Aikin2010,Brevik2010} & 1 & $3.2$ & $4.6$ & $0.6'$ & $2\%$ & \textbf{\textcolor[rgb]{0.50,0.50,0.00}{2.5}} \\
    \hline
    Planck (30 months) \cite{Ade:2011ah,Collaboration:2011ck}& 7 & $1.0 - 8.4$ & $1.7 - 14.5$ &  $4.7' - 32.7'$ & 80\% & \textbf{\textcolor[rgb]{0.00,0.40,0.00}{5.9}} \\
    \hline
    SPT SZ \cite{Shirokoff:2010cs,Keisler:2011aw} & 3 & $5.7 - 30$ & $-$  & $1.0' - 1.6'$ & 6\% & \textbf{\textcolor[rgb]{0.50,0.50,0.00}{2.0}} \\
    \hline
    SPTPol (3 years) \cite{McMahon2009}  & 2 & $1.3 - 1.5$ & $1.9 - 2.1$ & $1.0' - 1.6'$ & 1.6\% & \textbf{\textcolor[rgb]{0.50,0.50,0.0}{2.5}} \\
    \hline
    SPTPol Wider (6 years) & 2 & $2.4 - 2.6$ & $3.3 - 3.7$ & $1.0' - 1.6'$ & 10\% & \textbf{\textcolor[rgb]{0.00,0.40,0.0}{5.2}} \\
    \hline
    ACTPol Deep (1 year) \cite{Niemack:2010wz} & 2 & $0.5 - 2.2$ & $0.7 - 3.1$ & $1.0' - 1.4'$ & 0.36\% & \textbf{\textcolor[rgb]{0.50,0.50,0.00}{1.4}} \\
    \hline
    ACTPol Wide (1 year) \cite{Niemack:2010wz} & 2 & $2.5 - 11$ & $3.5 - 16$ & $1.0' - 1.4'$ & 10\% & \textbf{\textcolor[rgb]{0.00,0.40,0.00}{4.4}} \\
    \hline
    ACTPol Wider (4 years) & 2 & $2.5 - 11$ & $3.5 - 16$ & $1.0'- 1.4'$ & 40\%  & \textbf{\textcolor[rgb]{0.00,0.40,0.00}{8.8}} \\
    \hline
    COrE (4 years) \cite{Collaboration:2011ck} & 15  & $0.07 - 9.0$ & $0.12 - 15.6$ & $2.8' - 23.3'$ & 80\% & \textbf{\textcolor[rgb]{0.00,0.40,0.00}{14}} \\
    \hline
    EPIC 4K \cite{Bock:2009xw}& 9 & $0.08 - 0.82$ & $0.11 - 1.2$ & $2.5' - 28'$ & 80\% & \textbf{\textcolor[rgb]{0.00,0.40,0.00}{16}}\\
    \hline
    EPIC 30K \cite{Bock:2009xw} & 9 & $0.20 - 4.4$ & $0.28 - 6.2$ & $2.5' - 28'$ & 80\%  & \textbf{\textcolor[rgb]{0.00,0.40,0.00}{13}} \\
    \hline
    $\,$Ideal Exp.  (up to $\ell = 6000$)$\,$ & Any & $0$ & $0$ & $0'$ & 100\%  & \textbf{\textcolor[rgb]{0.00,0.40,0.00}{44}} \\
    \hline
\end{tabular}
\caption{Summary of CMB experiments. The second column relates the number of frequency channels observed; $\,\theta_{\rm fwhm}\,$ is the beam size diffraction limit with full width at half maximum; $\,\sigma_T\,$ is the thermodynamic temperature sensitivity per pixel; $\,\sigma_P\,$ likewise for the polarization quantities $E$ and $B$; $\,f_{\rm sky}$ indicates the fraction of the sky covered. The quoted ranges of sensitivity and resolution stand for the different frequency bands below 420 GHz (which we find to be the relevant ones in the cases here considered). In the last column we quote the computed signal-to-noise ratio, but one should note that these values depend somewhat on the fiducial spectra (see text). \label{tab:cmb-experiments}}
\end{table*}

If on the other hand one detects an aberration which is larger than the one expected due to $\,\delta\boldsymbol{\beta}_{\rm dip}$, this would imply that our original assumption (to wit, that the CMB temperature dipole is due only to a relative velocity, or to some other effect which is exactly degenerate with a velocity in all multipoles) is incorrect, and we could be measuring for the first time a primordial dipole contribution. The nature of such a contribution would have an impact on the high-$\ell$ correlations. If some type of perturbations are completely degenerate with a velocity effect also producing the same correlations at high-$\ell$, then they would be of course not detected by our method. In general since different type perturbations may lead to different predictions about the high-$\ell$ correlations, then this would lead to the possibility to distinguish for instance an adiabatic mode from an isocurvature dipolar perturbation or a  dipolar
lensing effect or some other exotic contribution such as one due to vector fields. We leave this important question for future work.

The idea of deboosting the CMB was first proposed in~\cite{Menzies:2004vr} with the goal of simplifying analysis of general CMB correlations. In the same paper it was pointed out that applying~\eqref{eq:deboosted-T} directly into the reconstructed temperature maps could lead to some propagation of errors (and perhaps generation of spurious correlations) arising from an interpolation of a pixelised map. The ideal solution is therefore to deboost the temperature field directly in the instrument raw time-ordered data (TOD). In other words, one should correct the direction and frequency of each point in the TOD by applying a $\,-\boldsymbol{\beta}_{\rm dip}^{\rm fit}\,$ deboost with $\,\boldsymbol{\hat{n}^\prime}\,$ corresponding to the precise pointing of the instrument (and orbital velocity, as the rotation of the Earth around the Sun introduces a $10\%$ modulation to the dipole) at the original time the photons were collected. The corresponding pre-deboosted \alm's must only be computed \emph{a posteriori} from the deboosted TOD.

\section{Applications to Current and Future CMB Experiments}\label{sec:applications}

\subsection{Summary of CMB Experiments}

In this Section we explore the idea in more detail and estimate the expected signal strength and detection possibilities in a number of CMB experiments. As it turns out, experiments which cover only a few percent of the sky are not favored to detect the proposed correlations, due to extra contribution from cosmic variance (see~\eqref{eq:totalnoise} and~\eqref{eq:fsky-correction} below). These include among others: CBI~\cite{Sievers:2009ah}, QUAD~\cite{Brown:2009uy}, QUIET~\cite{Bischoff:2010ud}, BICEP~\cite{Chiang:2009xsa}, SPIDER~\cite{Crill:2008rd,Filippini:2011ds,O'Dea:2011th}  and  POLARBEAR~\cite{Errard:2010bn}. We will focus instead on the experiments which cover a substantial fraction of the sky, in particular WMAP, Planck and SPT as well as some of the proposed future ones (SPIDER, ACTPol, SPTPol, Cosmic Origins Explorer -- COrE  and The Experimental Probe of Inflationary Cosmology -- EPIC). To illustrate the difficulties faced by surveys probing only a small piece of the sky we will make exceptions for some of those: ACBAR, EBEX and BICEP2.

Estimates in this section refer to statistical noise alone; care must be taken when interpreting these due to the presence of foregrounds and systematic noise.

In Tables~\ref{tab:cmb-experiments} and~\ref{tab:cmb-experiments-2} we list a summary of CMB experiments. In the former we list the range of some parameters on the different frequency bands; in the latter we consider only the most promising subset of experiments and moreover, for a more detailed comparison, consider only the best frequency channel for measuring the spectra. In both tables $\,\theta_{\rm fwhm}\,$ is the beam size diffraction limit with full width at half maximum (fwhm); $\,\sigma_T\,$ is the thermodynamic temperature sensitivity per pixel (a square the side of which is the fwhm extent of the beam); $\,\sigma_P\,$ likewise for the polarization quantities $Q$ and $U$; $\,\ell_{\rm cvlim}^T\,$ is the multipole at which the temperature noise spectrum equals cosmic variance; $\,\ell_{\rm max}^{T,E}\,$ is the multipole for which the $TT$ or $EE$ spectrum equals the instrument (statistical) noise (i.e., $S/N = 1$). For both SPTPol and ACTPol we assume a net 9-hour per day observation time.\footnote{This in principle could be made higher by making observations also during daytime, although this extra exposure would be done with less sensitivity due to heating of the telescope~\cite{spergel-private}}

\begin{table*}[t]
\begin{tabular}{|c|c|c|c|c|c|c|c|c|}
    \hline
    \textbf{Experiment} & $\,$best $\nu \unit(GHz) \,$ & $\,10^6 \sigma_T (\frac{\mu K}{K})\,$ & $\,10^6 \sigma_{P}\,(\frac{\mu K}{K})\,$ & $\;\theta_{\rm fwhm}\;$ & $\;f_{\rm sky}\;$ & $\;\ell_{\rm cvlim}^T\;$ & $\;\ell_{\rm max}^T\;$ & $\;\ell_{\rm max}^E\;$ \\
    \hline \hline
    WMAP (9 years)  & 94 & $14$ & $20$ & $13.2'$ & 78\% & 600 & 900 & 5 \\
    \hline
    Planck (30 months)& 143 & $1.0$ & $1.7$ &  $7.2'$ & 80\% & 1800 & 2500 & 1700 \\
    \hline
    ACTPol Wider (4 years) & 150 & 2.5 & 3.5 & $1.4'$ & 40\% & 3600 & 4600 & 3300 \\
    \hline
    COrE (4 years) & 225  & $0.21$ & $0.36$ & $4.7'$ & 80\% & 3100 & 3700 & 3000 \\
    \hline
    EPIC 4K& 220 & 0.24  & 0.34 & $3.8'$ & 80\% & 3400 & 4300 & 3300 \\
    \hline
    EPIC 30K & 220 & 0.56 & 0.79 & $3.8'$ & 80\% & 3000 & 3900 & 3000 \\
    \hline
\end{tabular}
\caption{Similar to Table~\ref{tab:cmb-experiments} for selected CMB experiments which we will consider in more detail. Values quoted stand for the optimal frequency band with respect to a combination of temperature and angular sensitivities. $\,\sigma_T\,$ is the thermodynamic temperature sensitivity per pixel; $\,\sigma_P\,$ likewise for the polarization quantities $E$ and $B$; $\,\ell_{\rm cvlim}^T\,$ is the approximate multipole at which the temperature noise spectrum equals cosmic variance; $\,\ell_{\rm max}^{T,E}\,$ is the approximate  multipole for which the $TT$ or $EE$ spectrum equals the instrument (statistical) noise. Note that $\,\sigma_T\,$, $\,\theta_{\rm fwhm}\,$, $f_{\rm sky}$, $\,\ell_{\rm cvlim}^T\,$  and $\,\ell_{\rm max}^X\,$  are related by equation~\eqref{eq:noise}. \label{tab:cmb-experiments-2}}
\end{table*}

The quantities $\,\theta_{\rm fwhm}\,$, $\,\ell_{\rm cvlim}^T\,$, $f_{\rm sky}$ and $\,\sigma_T\,$ are related by the expression for the noise power-spectrum (see \textit{e.g.}~\cite{Dodelson:2003ft}):
\begin{equation}
      \Delta C_{\ell} =  \sqrt{\frac{2}{f_{\rm sky}(2\ell + 1)}} \, \Big[ C_\ell \,+\, N_\ell \Big] ,\label{eq:totalnoise}
\end{equation}
where the first term stands for the cosmic variance ($CV_{\ell}$) and $N_\ell$ for the instrumental noise (see~\cite{Knox:1995dq,Dodelson:2003ft,Perotto:2006rj,Nolta:2008ih}):
\begin{align}
      N_{\ell} \;=\; \theta_{\rm fwhm}^{2}\sigma_{T}^{2} \,\exp\left[\ell(\ell+1)\frac{\theta_{\rm fwhm}^{2}}{8\ln{2}}\right] .\label{eq:noise}
\end{align}
Cosmic variance is predominant at lower and intermediate values of $\ell$ (to wit for $\ell < \ell_{\rm cvlim}^T$); instrumental noise dominates for $\ell > \ell_{\rm cvlim}^T$ and determines $\ell_{\rm max}^X$. Note that in the case of Earth experiments the estimate~\eqref{eq:noise}, sometimes referred to as the Knox formula, is inaccurate for $\,\ell \lesssim 500\,$ due to possible atmospheric fluctuations~\cite{Niemack:2010wz}. Note also that we list sensitivities per pixel in $\mu K / K$, but some of the references here listed prefer to describe them as either noise-equivalent temperatures (NET) in $\,\mu K_{\rm CMB} \sqrt{s}\;$ (for a single detector -- e.g. one bolometer) or as sensitivity in $\,\mu K \cdot\,$arcmin (for a given frequency band, with all detectors in that band combined). To convert between these quantities one has to make use of the following relations~\cite{Bock:2009xw}:
\begin{align}
      \sigma (\mu K \cdot \mathrm{arcmin}) &\,=\, \sqrt{\frac{8\pi \tilde{f}_{\rm sky} \big[\mathrm{NET}(\mu K_{\rm CMB} \sqrt{s})\big]^2}{ t_{\rm mission}(s) \; N_{\rm detectors}}} \frac{10800}{\pi} \,, \label{eq:NET-muarcmin} \\
      \sigma \left(\frac{\mu K}{K}\right) &\,=\, \frac{\sigma (\mu K \cdot \mathrm{arcmin})}{2.725K \; \theta_{\rm fwhm}(\mathrm{arcmin})} \, . \label{eq:muarcmin-muKoverK}
\end{align}
Note that in~\eqref{eq:NET-muarcmin} above, $\tilde{f}_{\rm sky}$ is the total fraction of the sky covered by the corresponding CMB mission \emph{before} any possible sky cuts (usually around the galactic plane). Thus, $\,\tilde{f}_{\rm sky} = 1\,$ for all space based surveys and we assume $\,\tilde{f}_{\rm sky} = f_{\rm sky}\,$ for all ground based surveys here considered.

Also worth noticing is the relation between $\sigma_{T}$ and $\sigma_{E;B}$. If both linear polarization states are given equal integration times (as is usually the case), the total number of photons available for the temperature measurement will be twice the number available for either polarization measurement, and one has $\,\sigma_{E;B} = \sqrt{2} \sigma_{T}$~\cite{Kamionkowski:1996ks}. This relation, however, assumes that all detectors (being either coherent receivers, like HEMT amplifiers, or incoherent ones, like bolometers) in the experiment are sensitive to polarization. If that is not the case, then one has instead the relation
\begin{equation}
    \sigma_{E;B} \,=\,  \sqrt{2\frac{N_{\rm total}}{N_{\rm polar}}} \, \sigma_{T}  ,\label{eq:noise-relation}
\end{equation}
where $\,N_{\rm total}\,$ stands for the total number of detectors and $\,N_{\rm pol} \le N_{\rm total}\,$ is the number of polarization detectors.

Finally, the measurements in the different frequency bands are formally independent, so one can combine different channels to get lower instrumental noises (cosmic variance is completely correlated in all frequency bands and therefore cannot be mitigated) and thus probe higher multipoles. For Planck, as an example, combining 2 channels (143 and 217 GHz) allows us to increase $\ell_{\rm max}^T$ by around 20\%. However, more important than an increase in $\ell_{\rm max}^T$, the lower instrumental noise after combining all channels means that the correlations here addressed can be detected with higher $S/N$ ratios. To wit, combining the different bands one gets
\begin{align}
      N_{\ell}^{\rm comb} &= \left[ \sum_{i=1}^{\# {\rm bands}} \Big(N_{\ell}^{i}\Big)^{-2} \right]^{-1/2}, \label{eq:noise-combined}
\end{align}
where $N_{\ell}^{i}$ is the instrument noise spectra for the $i$-th frequency band.

One must nevertheless be aware that different channels are subject to largely different foregrounds and systematics, so that actually the usable $f_{\rm sky}$ differs for each frequency range. In what follows, for simplicity, we will assume a fixed $f_{\rm sky}$ in all frequency bands for each experiment.

\subsection{Detectability of our Proper Motion}

In this subsection we seek to answer the following question: for a given CMB experiment, what is the smallest value of $\beta_{\rm res}$ it can in principle detect (say, $S/N = 1$ or 3)?

\begin{figure*}[t]
    \includegraphics[width=8.4cm]{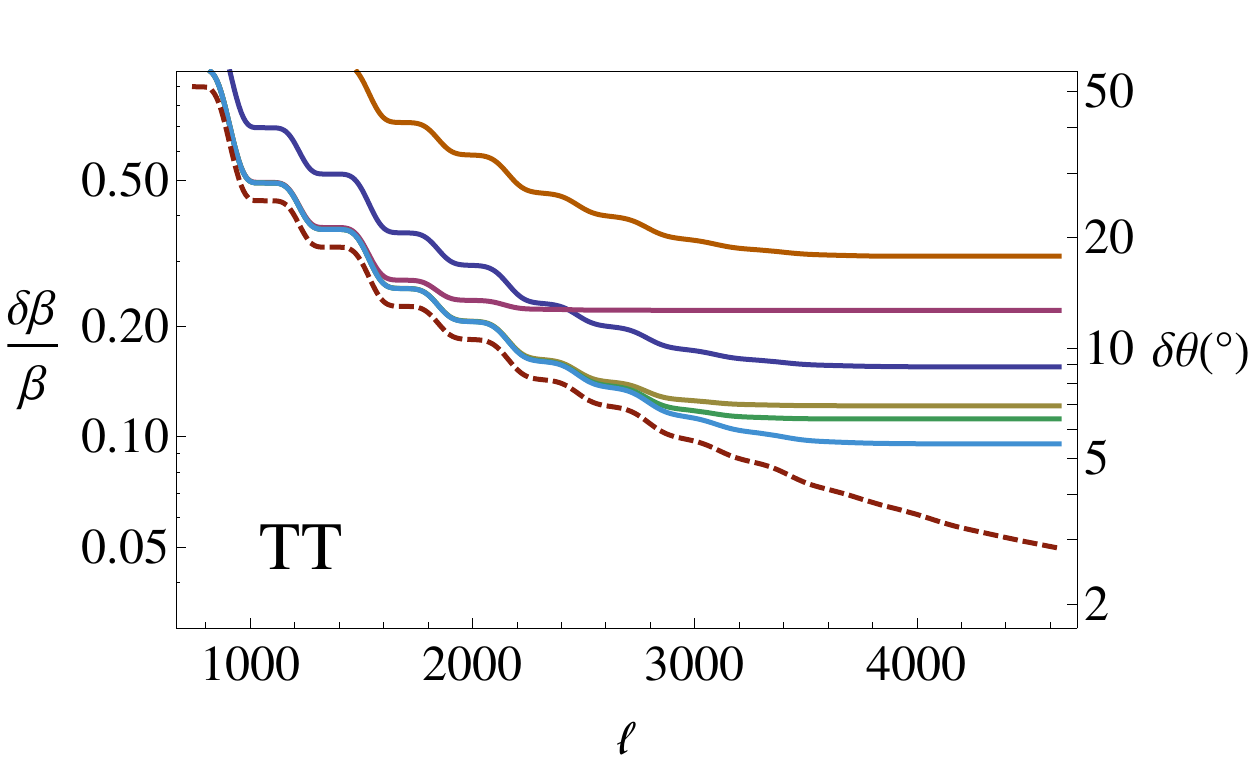}\qquad
    \includegraphics[width=8.4cm]{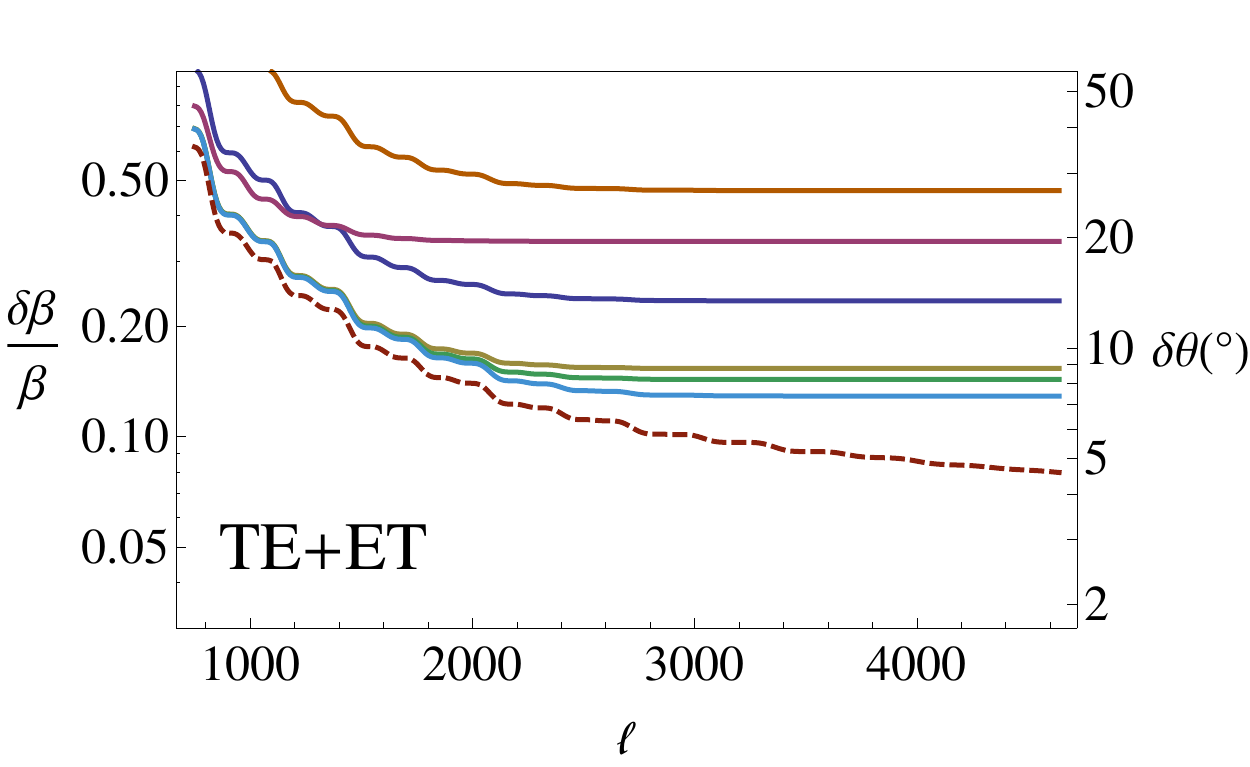}
    \includegraphics[width=8.4cm]{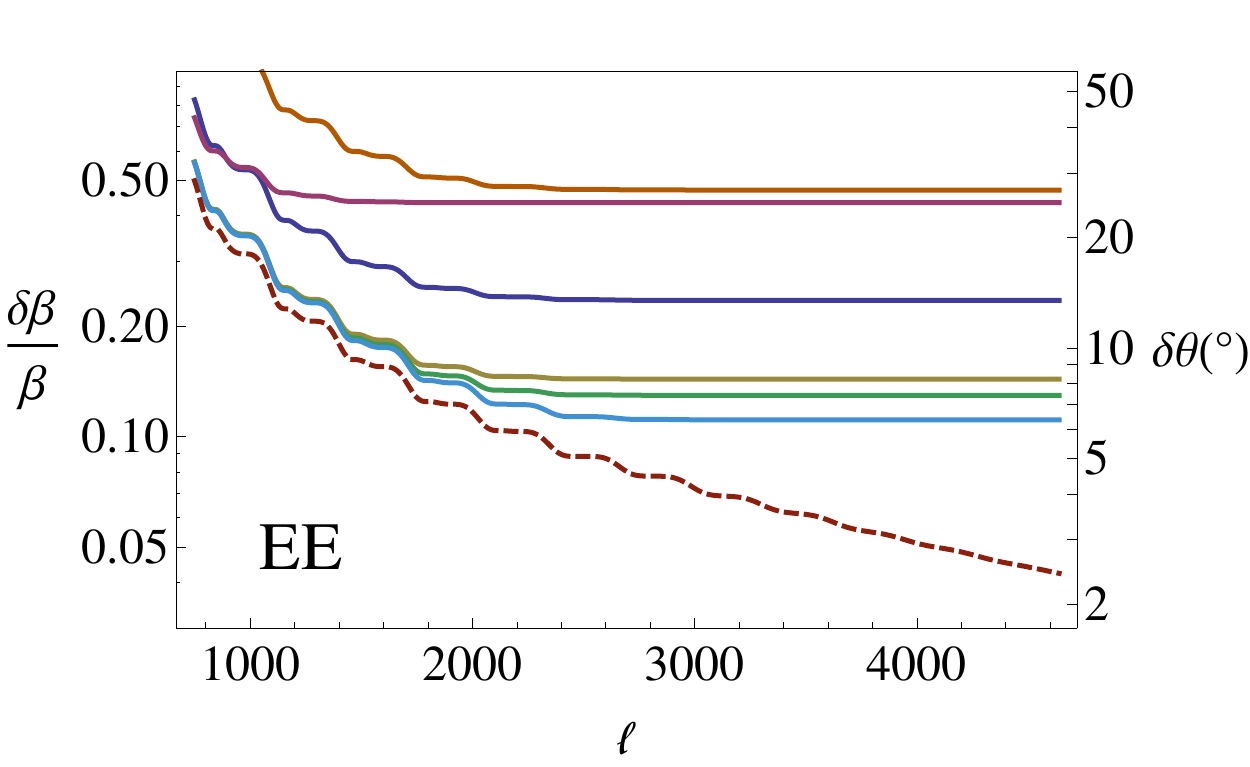}\qquad
    \includegraphics[width=8.4cm]{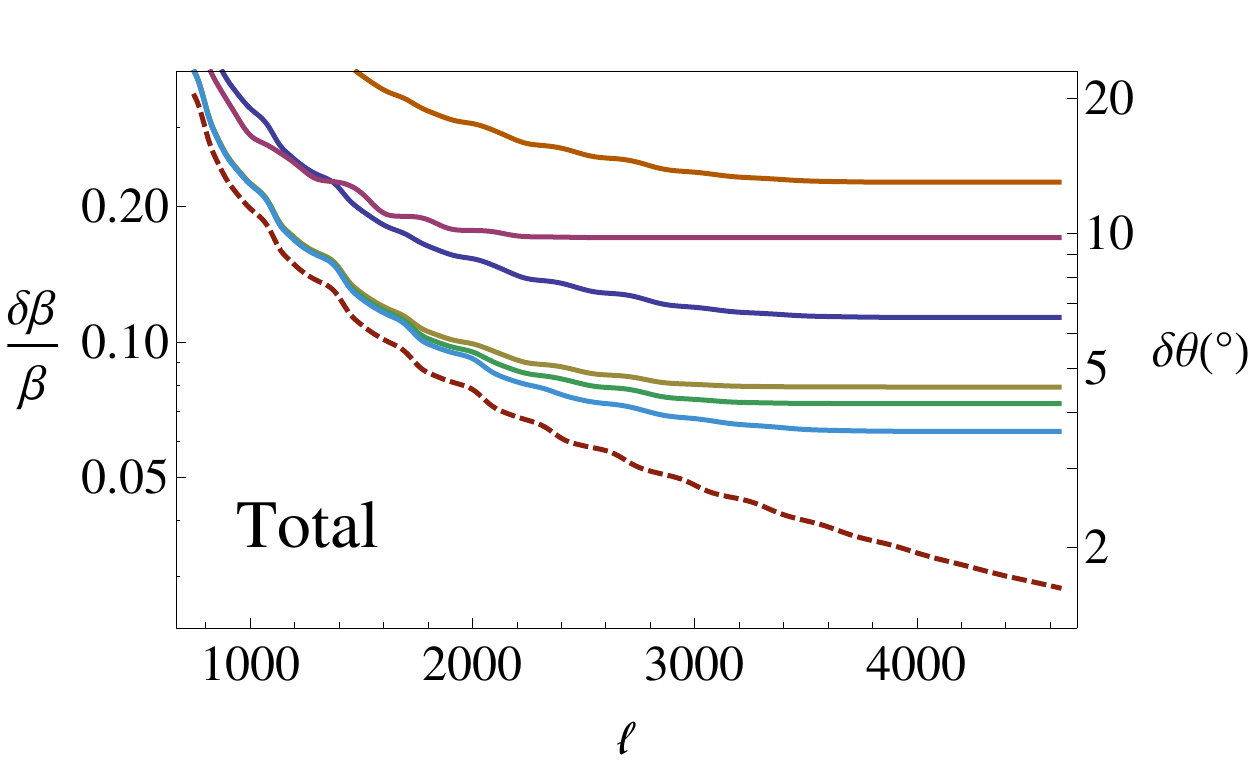}
    \caption{Precision in the measurement of $\delta \beta$ achieved by different experiments using the temperature $T$, the $E$ mode of polarization the $TE$ cross power-spectra and the combination of all these. Here we still assume $\beta_{\rm res} = 1.231 \times 10^{-3}$ (i.e., no pre-deboosting). Top to bottom: ACTPol Wide (1 year); Planck (30 months);  ACTPol Wider (4 years); EPIC (30K); Core (4 years); EPIC (4K) and finally, represented by the dashed line, an ideal experiment (no noise).}
    \label{fig:deltabeta}
\end{figure*}

Using the induced off-diagonal $a_{\ell m}$ correlations, an estimator for our peculiar velocity was built in~\cite{Amendola:2010ty}. We summarize below the procedure. We first define the following basic quantities:
\begin{equation}
    F_{\ell m}^{XY}\equiv a_{\ell\, m}^{X*}a_{\ell+1\, m}^{Y}\,,
\end{equation}
in the appropriate frame,
where $X,Y$ stands for either temperature ($T$) or one of the polarization channels ($E,B$). The useful quantities are the real part\footnote{The imaginary part has zero average. Note however that in the presence of some primordial parity violating effect the imaginary parts of $F^{EB}_{\ell m}$ and $F^{TB}_{\ell m}$ could be be nonzero, already in the CMB rest frame. Such observables would presumably be small and the correction due to a boost would have a further suppression due to $\beta$, so this would most likely be a negligible correction at large scales. However such correction due to a boost may be relevant at very small scales when $\beta \ell \sim 1$. Investigating in detail such possibility can be interesting in a specific parity violating scenario, which goes beyond the scope of our present work.}  of the above
\begin{equation}
    f_{\ell m}^{XY}\equiv \frac{1}{2}(F_{\ell m}^{XY}+F_{\ell-m}^{XY})\,.
\end{equation}
Note that $f_{\ell m}^{XY} = f_{\ell -m}^{XY}$. Given a peculiar velocity $\beta_{\rm res}$, we can predict the average value of the $f_{\ell m}^{XY}{}'s$, which are given by~\cite{Amendola:2010ty}:
\begin{equation}
    \langle f_{\ell m}^{XY}\rangle \;=\; c_{\ell+1m}^{-Y}C_{\ell}^{XY}+c_{\ell m}^{+X}C_{\ell+1}^{XY}\,,\label{eq:fth}
\end{equation}
The $c_{\ell m}^{\pm X}$ coefficients are different in the case of temperature and polarization and are given by~\eqref{eq:non-linear-fit-n1}; to wit:
\begin{equation} \label{eq:clmT}
\begin{aligned}
    c_{\ell m}^{+X} & =  \beta_{\rm res}(\ell+1)\;{}_sG_{\ell+1\, m}\,, \\
    c_{\ell m}^{-X} & = -\beta_{\rm res}\,\ell \;{}_sG_{\ell\, m}\,,
\end{aligned}
\end{equation}
where $s=0$ for temperature and $s=2$ for polarization, and where we made use of~\eqref{eq:challinor-fun}. Note that in the limit of flat spectra ($C_{\ell}^{XY} = C_{\ell+1}^{XY}$), one has $\langle f_{\ell m}^{XY}\rangle = 0$. One can show, using~\eqref{eq:non-linear-fit-general}, that this remains true up to ${\cal O} \big( \beta^3 \big)$; we come back to this issue in the end of this Section. Therefore $\langle f_{\ell m}^{XY}\rangle$ is generically higher the more wiggled the power spectrum is.

Since $\langle f_{\ell m}^{XY}\rangle$ is proportional to $\beta_{\rm res}$, it is useful to define a related quantity
\begin{equation} \label{eq:flm-hat}
    \langle f_{\ell m}^{XY}\rangle \;\equiv\; \beta_{\rm res}\, \langle \hat{f}_{\ell m}^{XY}\rangle\,.
\end{equation}
From these predictions, we can compute an estimator
\begin{equation}
    \hat{\beta}=\left(\sum_{X,\ell,m}\frac{f_{\ell m}^{XY,\,{\rm obs}}\langle \hat{f}_{\ell m}^{XY}\rangle}{{\mathfrak C}_{\ell}{\mathfrak C}_{\ell+1}}\right)\left(\sum_{X,\ell,m}\frac{\langle \hat{f}_{\ell m}^{XY}\rangle^{2}}{{\mathfrak C}_{\ell}{\mathfrak C}_{\ell+1}}\right)^{-1},
\end{equation}
where (see~\cite{Amendola:2010ty,Perotto:2006rj,Verde:2009tu})\footnote{Note that in~\cite{Amendola:2010ty} there was a typo and the square root was missing in~\eqref{eq:fsky-correction}.}
\begin{equation}\label{eq:fsky-correction}
    {\mathfrak C}_{\ell} \,\equiv\, \frac{1}{\sqrt{f_{\rm sky}}} \big(C_\ell + N_\ell \big)\,
\end{equation}
and the sums are in principle carried out with $\,-\ell \le m\le \ell$, but can also be simplified to a sum on $0 \le m\le \ell$ since $f_{\ell m}^{XY} = f_{\ell -m}^{XY}$. An approximate value for the variance of $\hat{\beta}_{\rm res}$ can also be written as~\cite{Amendola:2010ty}
\begin{equation}
    \delta \beta \,\equiv\, \sqrt{\langle\hat{\beta}^{2}\rangle}\approx \left[\sum_{X}\sum_{\ell}\sum_{m=-\ell}^\ell \frac{\langle \hat{f}_{\ell m}^{XY}\rangle^{2}} {{\mathfrak C}_{\ell} {\mathfrak C}_{\ell+1}} \right]^{-\frac{1}{2}}=\, \beta_{\rm res} \left(\frac{N}{S}\right)\label{eq:delta-beta} .
\end{equation}
This leads to an important conclusion: the estimate of $\delta\beta$ does not depend on the value of $\beta_{\rm res}$ itself.  Note that this is physically motivated by the simple fact that the error in $\beta$ is mainly due to cosmic variance which, to leading order,  is independent on $\beta$. Therefore, one can determine the achievable precision of a given experiment without knowledge of the residual velocity $\beta_{\rm res}$. Likewise, one can also determine what is the minimal value $\beta_{\rm res}$ that a given experiment could detect (again, say, with $S/N = 1$  or 3).

Finally, it was also shown in~\cite{Amendola:2010ty} that the estimate on the direction of the velocity is directly related to the one of the magnitude. To wit if one chooses the fiducial velocity as the $z$-axis, for small angles the error on the absolute value of the velocity and the error in the magnitude of the direction are related by
\begin{equation}
    \delta\theta\,=\,\frac{\delta\beta}{\beta_{\rm res}}.
\end{equation}
Note that although as stated above the achievable precision in $\delta \beta$ does not depend on the fiducial value of the velocity (and thus does not depend whether one pre-deboosts the CMB or not), the same is not true for $\delta\theta$, so clearly the direction of the residual velocity will be measurable only if $\beta_{\rm res}$ is significantly larger than $\delta\beta$ for a given experiment.

\begin{figure}[t!]
    \includegraphics[width=8.1cm]{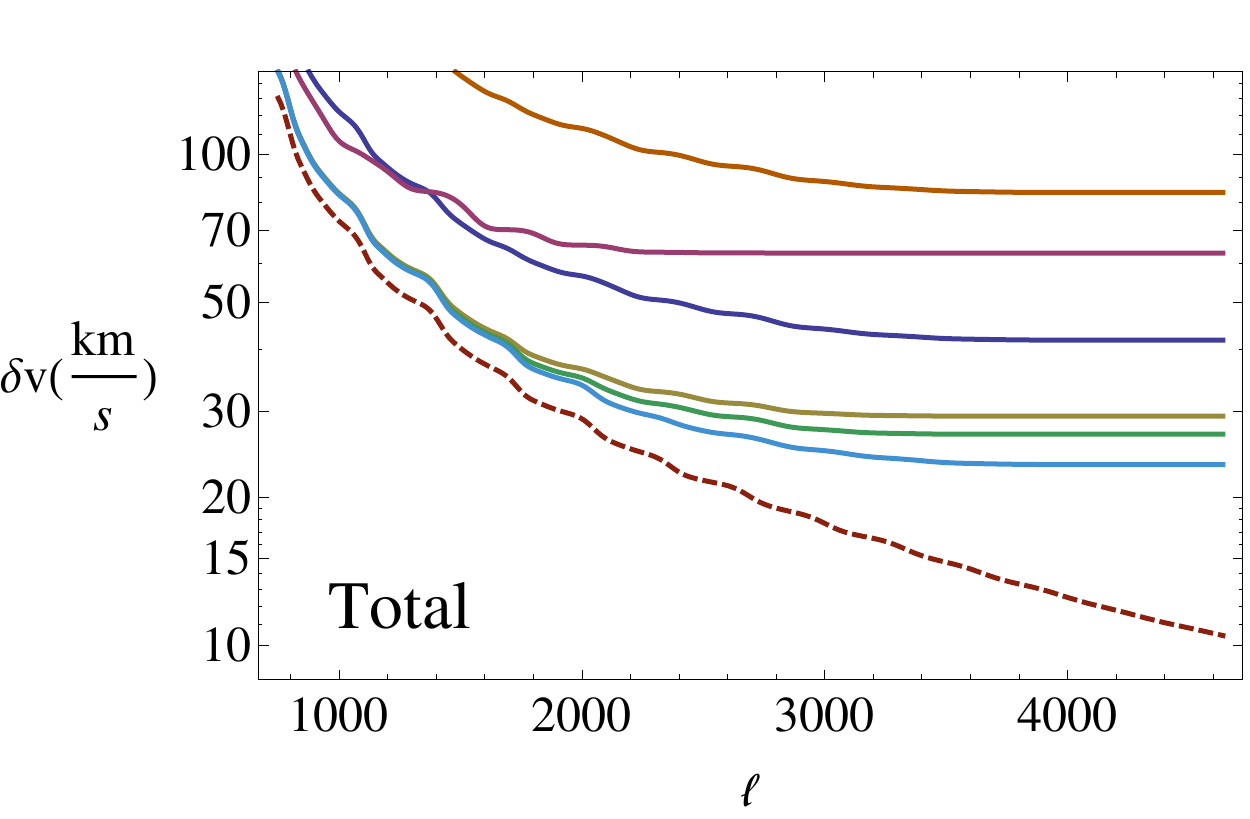}
    \caption{Similar as Figure~\ref{fig:deltabeta} but for the precision in the measurement of $\delta v$ achieved by different experiments using a combination of $TT$, $TE$ and $EE$ channels. Here we need not assume any value for $\beta_{\rm res}$. Top to bottom in both plots: ACTPol Wide (1 year);  Planck (30 months); ACTPol Wider (4 years); EPIC (30K); Core (4 years); EPIC (4K) and finally, represented by the dashed line, an ideal experiment (no noise).}
    \label{fig:delta-vel}
\end{figure}

Figure~\ref{fig:deltabeta} depicts the expected precision of some of the CMB experiments listed in Table~\ref{tab:cmb-experiments} as a function of the highest multipole $\ell$ taken into account. Note that these predictions depend somewhat on the assumed fiducial power spectra, so quoted values should not be taken at face value. Here we made use of CAMB~\cite{Lewis:1999bs} and took its default cosmological parameters as our fiducial ones (see Appendix~\ref{app:details-spectra} for more details). Experiments not represented in the figure cannot produce a detection since their maximum $S/N$ is below 1 by a good margin (except for WMAP, for which we forecast a $S/N \simeq 0.7$, which is almost borderline). Note that even the best experiments (Core and EPIC 4K) fall short of the (arguable) target of $\sim 1\%$ fractional precision on the velocity. Nonetheless, their precision would be able to detect a non-standard signal which amounts to at least $8\%$  in the CMB dipole. It will take an even better experiment, with sensitivities and beam sizes which are respectively high and small enough to probe temperature and polarization multipoles up to $\ell \simeq 5000$ to achieve the level of $2\%$. Note however that going to smaller and smaller scales the signal gets more and more contaminated by local sources, and if they are not taken properly into account this could prevent using this part of the spectrum, because it could introduce an additional preferred direction in the data.

In Figure~\ref{fig:delta-vel} we show a similar signal to the one of the bottom right plot of Figure~\ref{fig:deltabeta}, but in terms of absolute precision in the velocity, which does not depend on the value of $\beta_{\rm res}$ (and therefore, on whether one deboosts the CMB or not). The bottomline is that Planck could measure our peculiar velocity $v$ with a precision of $55 \unit{km/s}$, whereas Core and EPIC 4K/30K could do the same with only $20 - 25 \unit{km/s}$ of error. Finally, an ideal experiment probing temperature and polarization multipoles up to $\ell \simeq 5000$ could achieve $\delta v \simeq 8 \unit{km/s}$.

We have so far ignored the $B$-mode polarization channel. This is due to the fact that its spectra is supposed to be highly suppressed with respect to the $E$-mode one. In fact, assuming no tensor perturbation modes (and thus that the $BB$ spectra is due to gravitational lensing alone) we find that none of the proposed experiments here considered can detect the correlations with $S/N$ ratios larger than unity. To wit, we find that Core would have its $\,S/N \simeq 0.6\,$ while for Epic 4K we would have$\,S/N \simeq 0.8$. An ideal experiment could instead have for its $BB$ correlations a $S/N > 1$ if it went until $\ell>2000$, and $S/N \simeq 10$ if it went until $\ell=4000$.

\begin{figure}[t]
    \includegraphics[width=8.4cm]{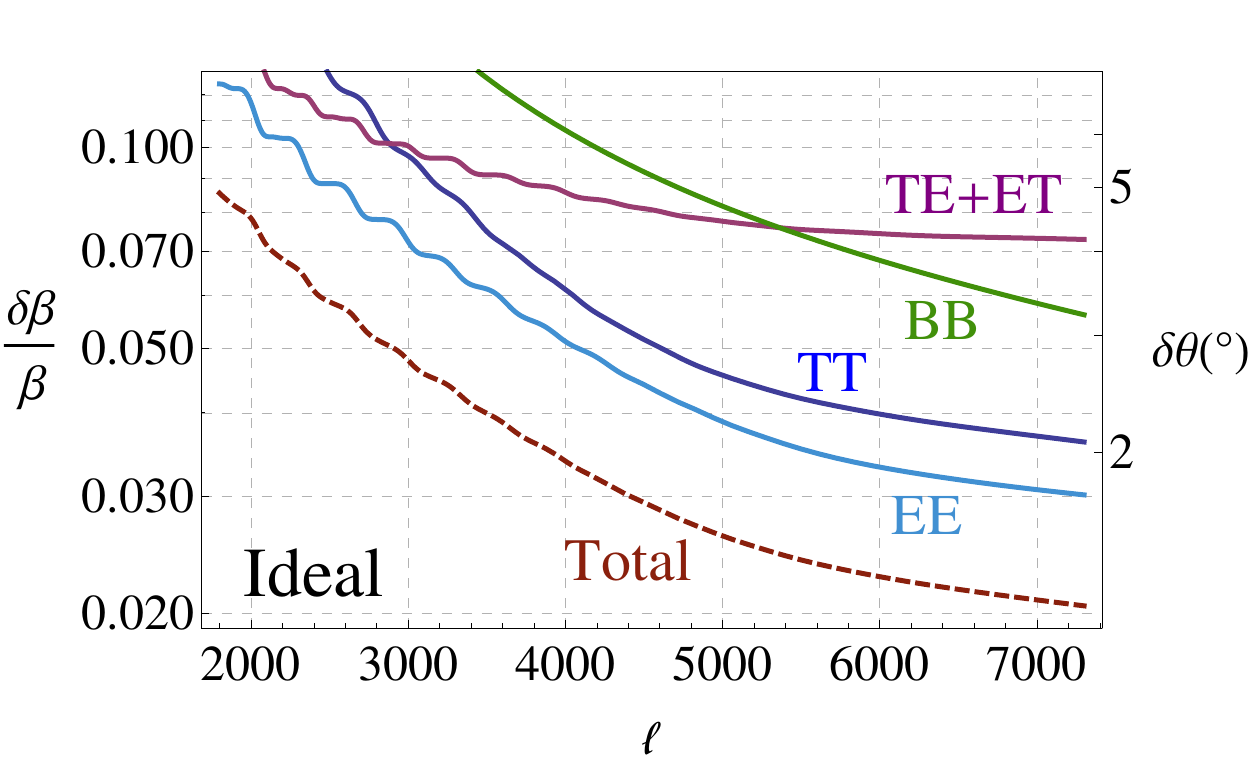}
    \caption{Maximum theoretical accuracy a CMB experiment can achieve in measuring the correlations in the different channels. Note that unlike all proposed CMB experiments here analyzed, correlations in the $BB$ channel are relevant here, and in fact its $S/N$ surpasses the $ET+TE$ case.}
    \label{fig:delta-beta-ideal-lensed}
\end{figure}

Figure~\ref{fig:delta-beta-ideal-lensed} depicts the maximum theoretical accuracy a CMB experiment can achieve in measuring the correlations in the different channels. Note that unlike all proposed CMB experiments here analyzed, correlations in the $BB$ channel are relevant here, and in fact its $S/N$ surpasses the $ET+TE$ case. One of the reasons for this is the fact that for very high scales ($\ell > 5000$) the $BB$ spectra is not too much smaller than the $EE$ and $TE$ ones (see Figure~\ref{fig:lensed-spectra}).

\begin{figure}[t]
    \includegraphics[width=8.2cm]{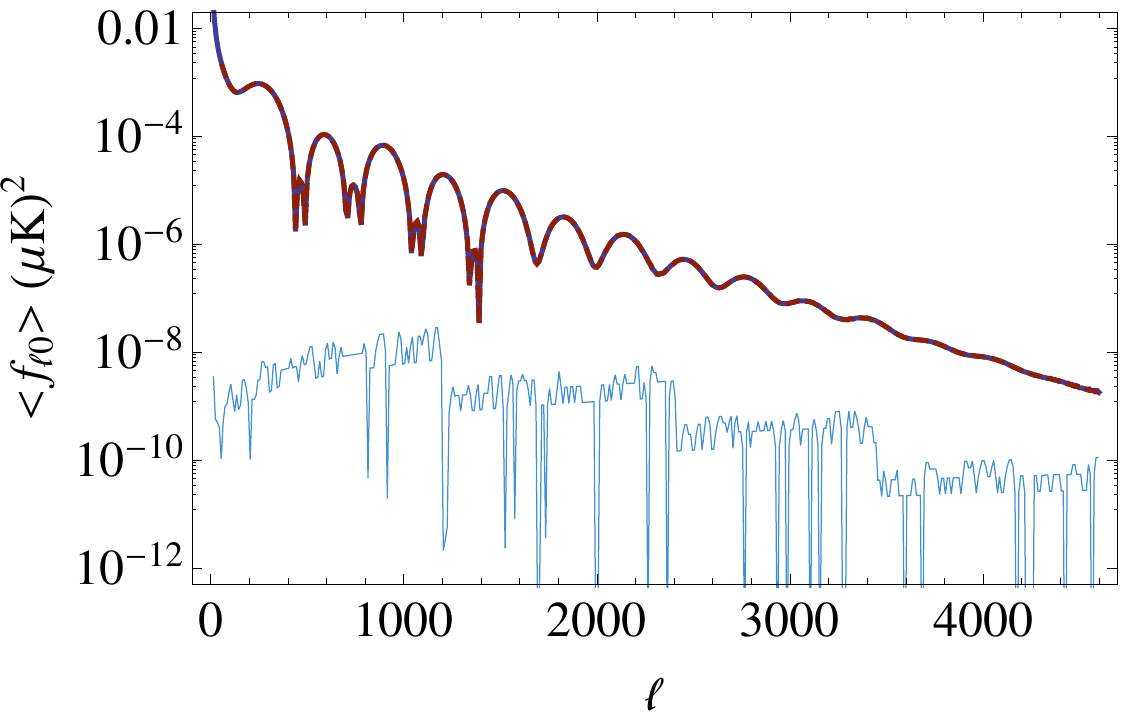}
    \caption{Comparison between $\left<f_{\ell m}^{TT}\right>$ for $m=0$ evaluated to first and up to third order (upper two curves, superimposed) and assuming a very high $\,\beta_{\rm res} = 5 \times 10^{-4}$. The difference between both curves is depicted below (light blue curve). We smoothed out this last curve (as it oscillates heavily around zero) for better clarity. Note that even for such a high residual velocity the error is only $\sim 1\%$ and that for $|m|>0$ the relative error is even smaller. }
    \label{fig:flm-3rd order}
\end{figure}

The results in Section~\ref{sec:nonlinear} allow us to estimate the accuracy of the first order expansion we use here, as $\beta_{\rm res} \ell$ approaches unity. A direct estimate would be obtained by comparing the non-linear fitting functions~\eqref{eq:non-linear-fit-n1} with a linear expansion. However, a more careful comparison has to be done at the level of the $\left<f_{\ell m}\right>$, because of the leading-order cancellations, which could be disrupted by small changes in the coefficients. We therefore evaluated, using~\eqref{eq:non-linear-fit-general}, $\left<f_{\ell m}\right>$ up to third order in $\beta$ (second order contributions are identically zero for the $\{\ell,\,\ell+1\}$ correlations) and compared the result with the first order $\left<f_{\ell m}\right>$. Amazingly, the leading order cancelation is basically undisturbed, and the correction to $\left<f_{\ell m}\right>$ is found to be small even for high values of $\beta_{\rm res}$: it is $\sim 1\%$ for $\beta_{\rm res} = 5 \times 10^{-4}$.  Figure~\ref{fig:flm-3rd order} depicts this property of $\left<f_{\ell m}\right>$.


\section{Conclusions}

In this work we proposed a method to overcome the challenges posed to the measurement of our peculiar velocity $\boldsymbol{\beta}$ through its aberration and Doppler  effect in the CMB at very small scales, due to the breakdown of the perturbative calculations when $\ell \gg 1/\beta$. This technique consists of pre-deboosting the CMB (at the level of the TOD) and subsequently evaluate the estimators originally constructed in~\cite{Amendola:2010ty}, which are based on a linear order Taylor expansion in $\beta$. The expected small residual velocity after such deboost justifies the use of only the linear order terms, and greatly simplifies the analysis. In particular, the pre-deboost method validates analysis of the aberration effect constrained to the correlations between $\{\ell,\,\ell+1\}$ only.

Making use of the estimator of the velocity derived in~\cite{Amendola:2010ty} for such correlations, we investigated the precision with which many of the current and future CMB experiments might be able to measure our velocity $v$. We find that Planck will put an error bar $\,\delta v \simeq 60$ km/s, similar to what ACTPol could achieve in only 2 years with a survey covering $40\%$ of the sky (4 years of observation would result in $\,\delta v \simeq 40$ km/s). SPTPol, could also be competitive if it carried out a similar wide survey. Proposed future space experiments such as EPIC and Core could put a limit $\,\delta v < 30$ km/s. Even more precise experiments, able to measure the CMB signals (and get systematics under control) all the way to $\ell \simeq 5000$ could in principle achieve  $\,\delta v < 10$ km/s.

Since one expects from the CMB dipole that $v \simeq 370$~km/s, with current (near future) experiments one would be able to detect a residual term of primordial origin which contributes to a fraction larger that $20\%$ ($10\%$) of the dipole. This is of course true unless the primordial effect itself also induces a dipolar kernel leading to correlations, with exactly the same coefficients as the peculiar velocity we analyze here, on the $\ell>1$ multipoles. Apart from velocity, other possible components to the dipole and/or a dipolar kernel include intrinsic adiabatic, isocurvature and lensing effects, as well as more exotic effects, some of which we now briefly discuss.

An example would be given by some fundamental vector field or perhaps magnetic fields coherent over the horizon scale, which could single out a preferred direction and therefore induce a dipole. Some inhomogeneous and anisotropic cosmological models are also expected to contribute to the dipole. For instance inhomogeneous gigaparsec-scale void models, embodied by the Lema\^itre-Tolman-Bondi metric, are sometimes constructed to serve as a candidate for dark energy, and have recently drawn considerable attention from the literature (see, for instance,~\cite{Marra:2011ct}). In these spherical symmetric models, the observer is sometimes assumed to be in the center for simplicity, but any off-center displacement induces a dipole contribution. This in fact happens to be the tightest constrain in this off-center distance, if one assumes the CMB dipole to be entirely due to this displacement and that any additional peculiar velocity is actually zero~\cite{Quartin:2009xr}. Metrics which exhibit vorticity (e.g., Bianchi V and VII) also can contain a preferred direction and sense, and so also introduce a dipole~\cite{Barrow1985}. On the other hand, homogenous anisotropic metrics without vorticity, such as Kantowsky-Sachs, Bianchi I and III, exhibit a preferred direction but no preferred sense (i.e., a preferred axis, but not an arrow), so they contribute to the CMB angular power spectrum at most at the quadrupole level~\cite{Graham:2010hh,Koivisto:2010dr}. Finally, it is also possible that more generically large-scale vector perturbations in FLRW with unusually large amplitude may induce such dipoles, as well as a primordial power spectrum which explicitly depends on the direction of the Fourier mode $\mathbf{k}$ instead of just $k=|\mathbf{k}|$~\cite{Hanson:2009gu}.



It is an interesting question, which deserves future investigation,  to check what kind of primordial perturbations or other effects such as the ones discussed above could induce similar correlations between different $\ell$'s and if the produced correlations are proportional to the produced Doppler effect. Since it is natural to expect that in some cases there is no such proportionality, aberration can provide in principle an observational handle to distinguish between these different contributions.

While the pre-deboost is probably the simplest way to proceeed, we have also proposed very precise fitting functions for the complicated integrals~\eqref{eq:non-linear-coef}, valid also when $\ell \gg 1/\beta $.
Although another method was recently proposed in~\cite{Chluba:2011zh} to compute this kernel elements, it relies on recursion relations between some of its elements and is not as simple to implement as just using the simple Bessel functions~\eqref{eq:non-linear-fit-general} and~\eqref{eq:non-linear-fit-n0}.


Apart from introducing a correlation between different \alm 's, it was showed in~\cite{Challinor:2002zh} that our peculiar velocity would also cause a bias in the measured CMB \emph{intensity} power spectrum proportional to $4\beta^2$. Here we showed instead that for the \emph{temperature} power spectrum (the one which is physically of more interest) the bias is exactly \emph{zero} at ${\cal O}\big(\beta^2\big)$.
We also showed that the small cross-correlation between $E$ and $B$ modes computed in~\cite{Challinor:2002zh} is instead also \emph{zero} -- at least up to ${\cal O}\big(\beta^6\big)$ -- when one corrects the aberration kernel to transform as temperature instead of as intensity.

With some recent claims of the existence of possible unexpected bulk flows~\cite{Watkins:2008hf,Kashlinsky:2008ut,Kashlinsky:2009dw,AtrioBarandela:2010wy} it is a very interesting prospect that we will soon have a complementary measurement of our peculiar velocity with respect to the CMB. A dipolar effect has already been measured outside the CMB: in supernovae~\cite{Weyant:2011hs}, in the cosmic infrared background~\cite{Fixsen:2011qk} and also marginally in X-rays~\cite{Boughn:2002bs}; it could also be measured in the future using the cosmic parallax technique~\cite{Quercellini:2008ty,Quartin:2009xr,Quercellini:2010zr}, and in the optical, radio and gamma bands.
It would therefore be interesting to explore the interconnections and compare the future precision that these different approaches can achieve.

\section*{Acknowledgments}

We would like to thank Fernando Atrio-Barandela, Carlo Baccigalupi, Tom Crawford, Thiago Pereira, David Spergel and Alex Wuensche for input which was useful in the development of this work. Special thanks to Luca Amendola and Riccardo Catena for useful discussions into the main idea behind this project and for work on related projects.

MQ is grateful to Brazilian research agency CNPq and the Research Council of Norway for support and to ITP, Universität Heidelberg for hospitality during the development of this project.

\appendix

\section{Details in the Computation of the Non-linear Coefficients}\label{app:details-numerical}

The integrand in~\eqref{eq:non-linear-coef} is an oscillating one, and care must be taken in the computation of the integral. The frequency of oscillations increase with $\ell$ (more precisely with $\ell - |m|$), which means that for smaller scales the numerical challenges increase. In fact, for $\ell > 8$ some standard numerical integrating procedures start to give incorrect results. In order to circumvent this one must resort to arbitrary-precision arithmetic and make use of higher-than-double precision (for an alternative method see~\cite{Chluba:2011zh}). With arbitrary-precision arithmetic we were able to compute some of the coefficients~\eqref{eq:non-linear-coef} but as $\ell$ increases so must the numerical precision, and the code both gets slower and memory used gets higher exponentially with precision. The smallest scale integral we computed, $\ell = 700$, required a working precision of 300 digits, took around half an hour to run in a desktop computer for a single value of $\,m\,$ and required roughly 5GB of RAM using Mathematica. In contrast, one $\ell = 200$ integral can be safely computed with 120 digits and take less than 5 minutes in the same machine.

\begin{figure}[t!]
    \includegraphics[width=7.8cm]{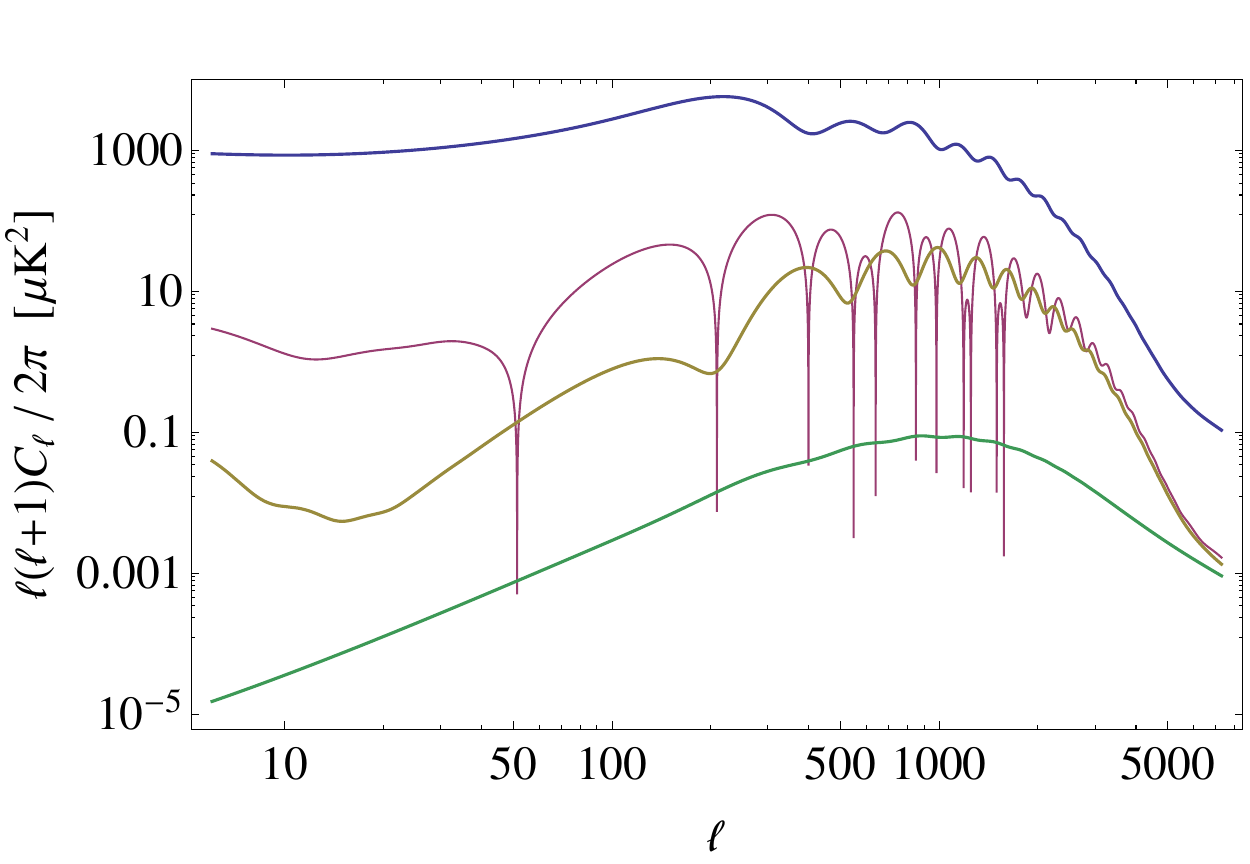}
    \caption{Computed angular power spectra with the parameters here listed and inclusion of lensing effects due to the linear part of the matter power spectrum. Top to bottom: $TT$, $TE$ (absolute value), $EE$ and $BB$ modes. Note that for very high $\ell$ the $BB$ spectrum approaches the $EE$ and $TE$ ones}
    \label{fig:lensed-spectra}
\end{figure}

\section{Details in the Computation of the Fiducial Spectra}\label{app:details-spectra}

The angular power spectra used here were obtained using the CMB code CAMB~\cite{Lewis:1999bs}, January 2011 release. As mentioned in the text, the predicted $S/N$ ratios here quoted depend somewhat on these fiducial spectra. These in turn depend not only on the fiducial cosmological parameters used, but also on the numerical accuracy of the code on small scales. For the experiments here considered the smallest scales are roughly $\,\ell \sim 4000\,$ but we were interested also in investigating scales all the way to $\,\ell \sim 7000\,$ for the case of an ideal experiment. To get reliable results in these scales we set in CAMB the \emph{Sample Boost} and both \emph{Accuracy Boost} parameters to 5 (setting some of these too high seemed to produce numerical artifacts for $\ell \gtrsim 6000$, such as negative $EE$ spectrum at $\ell \simeq 7500$), the \emph{Accurate $EE$ / $BB$} flags to true and used $\ell_{\rm max} = 11000$, $k\eta_{\rm max} = 44000$ for both scalar and tensor modes. The cosmological fiducial parameters were the default ones, such as: $h = 0.7$, $\Omega_b h^2 = 0.0226$, $\Omega_c h^2 = 0.114$, $\Omega_v = \Omega_k = 0$, $n=0.96$, $z_{\rm reion} = 11$ \emph{et cetera}. We also included the corrections due to gravitational lensing due only to the linear part of the matter power spectrum. It turns out that a lensed angular power spectrum leads to overall smaller $S/N$ in the correlations here studied, in particular in the polarization channels. The decrease in precision after summing all channels is roughly 20-30\% depending on the experiment. Figure~\ref{fig:lensed-spectra} depicts the generated spectra.


\bibliography{pre-deboost}

\end{document}